\documentclass[11pt,a4paper]{article} 
 \usepackage{jheppub}
\usepackage{enumerate}
\usepackage{epsfig} 
\usepackage{wasysym} 
\usepackage{mathrsfs} 
\usepackage{graphicx} 
\usepackage{amsfonts} 
\usepackage{amsbsy} 
\usepackage{amscd} 
\usepackage{pstricks} 
\usepackage{multirow} 
\usepackage{color}
\definecolor{myred}{rgb}{0.6,0,0} 
\definecolor{myblue}{rgb}{0,0.2,0.4}
\definecolor{mygreen}{rgb}{0,0.9,0.1}
\definecolor{hc}{rgb}{.9,0.1,0.7}
\definecolor{hcout}{rgb}{.9,0.7,0.9}
\definecolor{Orange}{rgb}{1.,0.65,0.}

\numberwithin{equation}{section}
\numberwithin{figure}{section}
\numberwithin{table}{section}

\newcommand{\be}{\begin{equation}}
\newcommand{\ee}{\end{equation}}
\newcommand{\bea}{\begin{eqnarray}}
\newcommand{\eea}{\end{eqnarray}}

\title{Left-Right  Symmetry \\ and the Charged Higgs Bosons at the LHC} 

\author[a]{G. Bambhaniya,}
\author[b]{J. Chakrabortty,} 
\author[c]{J. Gluza,} 
\author[c]{M. Kordiaczy\'nska} 
\author[d]{and R. Szafron}
\affiliation[a]{Theoretical Physics Division, 
Physical Research Laboratory, \\
Navarangpura, Ahmedabad - 380009, India}
  \affiliation[b]{Department of Physics, Indian Institute of Technology, Kanpur-208016, India} 
  \affiliation[c]{Institute of Physics, 
    University of Silesia, Uniwersytecka 4, PL-40-007 Katowice,
    Poland}
   \affiliation[d]{Department of Physics, University of Alberta, Edmonton, AB T6G 2E1, Canada}
\emailAdd{gulab@prl.res.in}
\emailAdd{joydeep@iitk.ac.in}
\emailAdd{gluza@us.edu.pl}
\emailAdd{mkordiaczynska@us.edu.pl}
\emailAdd{szafron@ualberta.ca}

\abstract{
The charged Higgs boson sector of the Minimal Manifest Left-Right Symmetric 
model (MLRSM) is investigated in the context of LHC discovery search for new physics beyond Standard Model.
We discuss and summarise the main processes within MLRSM  where heavy charged   Higgs bosons can be produced at the LHC. We explore the sce\-na\-rios where the amplified signals due to relatively light charged scalars dominate  against heavy neutral $Z_2$ and charged gauge $W_2$ as well as heavy neutral Higgs bosons signals which are dumped due to large vacuum expectation value $v_R$ of the right-handed scalar
triplet. Consistency with FCNC effects implies masses of two neutral Higgs bosons $A_1^0,H_1^0$ to be at least of 10 TeV order, which in turn implies that in MLRSM only three of four charged Higgs bosons, namely $H_{1,2}^{\pm \pm}$ and $H_1^{\pm}$,  can be simultaneously light. In particular, production processes with one and two doubly charged
Higgs bosons are considered. We further incorporate the decays of those scalars leading to multi lepton signals at the LHC.
Branching ratios for heavy neutrino $N_R$, $W_2$ and $Z_2$ decay into charged Higgs bosons are calculated. These effects are substantial enough and cannot be neglected.
The tri- and four-lepton final states for different benchmark points are analysed. 
Kinematic cuts are chosen in order to strength the leptonic signals and decrease the Standard Model (SM) background.
The results are presented using di-lepton invariant mass and lepton-lepton separation 
distributions for the same sign (SSDL) and opposite sign (OSDL) di-leptons as well as the charge asymmetry are also discussed.
 We have found that for considered MLRSM processes tri-lepton and four-lepton signals are most important for their detection when compared to the SM background. Both of the signals can be detected at 14 TeV collisions at the LHC with integrated luminosity at the level of $300~fb^{-1}$ with doubly charged Higgs bosons up to approximately 600 GeV.
Finally, possible extra contribution of the charged MLRSM scalar particles to the measured Higgs to di-photon ($H_0^0 \to \gamma \gamma$) decay
is computed and pointed out. 
}

\preprint{LPN13-089, Alberta Thy  7-13}

 
\begin{document}
\maketitle

\section{Introduction}

The LHC machine is working incredibly well shifting up the discovery limits for all the non-standard masses. For the same reason it is also true 
for the non-standard couplings and their possible values are shrinking more and more. Good examples are parameters 
connected with Left-Right (LR) symmetric models. These models enjoy richness of several types of beyond-the-SM particles \cite{Mohapatra:1974gc,Senjanovic:1975rk}. 
No wonder that these models are interesting for theoretical and phenomenological studies, for some recent works see \cite{Nemevsek:2011hz,Ferrari:2000sp,Frank:2011rb,Das:2012ii,Chakrabortty:2012pp,Dev:2013oxa,He:2012zp}   
and explored also by the LHC collaborations.

The searches at CMS and ATLAS have tightened up the limits on the masses of heavy gauge bosons.
Let us mention that before the LHC era the fits
to low energy charged and neutral currents were quite modest, e.g. for a charged gauge boson PDG reports $M_{W_2} > 715$ GeV \cite{Beringer:1900zz, Czakon:1999ga}. The new LHC analysis pushed the limits already much above 2 TeV \cite{CMS:kxa, CMS:2013qca, ATLAS:2012qjz, TheATLAScollaboration:2013iha, ATLAS:2013jma, CMS:2012uaa, salo:1558322}.
All these searches provide robust bounds on the extra gauge bosons,
for instance, the present limit for a charged heavy boson coming from the 
``golden" decay chain $W_R \to l_1 N_l \to l_1 l_2 jj$ is \cite{CMS:2012uaa,salo:1558322}

\begin{eqnarray}
M_{W_2} &\geq & 2.8 \; {\rm TeV}.
\label{mw2lhc}
\end{eqnarray}

This limit (at 95 \% C.L.) is for a genuine left-right symmetric model which we consider here (MLRSM)  with $g_L = g_R$ and three degenerate generations of heavy neutrinos and it is based on $\sqrt{s}=8$ TeV data. Typically, also limits for $Z_2$ mass are already beyond 2 TeV.    
 
The combined LEP lower limit on the singly charged Higgs boson mass is about 90 GeV \cite{Searches:2001ac}.
At the LHC, established  limits for singly charged Higgs boson masses  are 
\begin{eqnarray}
M_{H^\pm} & = & 80 \div 160 \;{\rm GeV},
\label{mhplimit}
\end{eqnarray}
if $BR(t \to H^+ b)< 5\%$ \cite{Chatrchyan:2012vca} and for higher masses than 160 GeV, see the limits in \cite{ATLAS-CONF-2013-090}.

For doubly charged Higgs bosons the analysis gives lower mass limits in a range 
\begin{eqnarray}
M_{H^{\pm \pm}} & \geq & 445\; {\rm GeV} (409\; {\rm GeV})\;\;\; {\rm for} \;\;\; {\mbox{\rm  CMS (ATLAS)}},
\label{mhpplimit}
\end{eqnarray}
  in the 100\% branching fraction scenarios \cite{CMS:2012kua,ATLAS:2012hi}.

The mass limit for heavy neutrinos  is \cite{CMS:2012zv, Aad:2012dm}
\begin{equation}
M_{N_R} > 780\; {\rm GeV},
\end{equation}
 but it must be kept in mind that bounds on
$M_{N_R}$ and $M_{W_2}$ are not independent from each other \cite{CMS:2012uaa,salo:1558322}. 
Neutrinoless double beta decay allows for heavy neutrinos with relatively light masses,  see e.g. 
\cite{Mohapatra:1979ia,Mohapatra:1980yp,Maiezza:2010ic,Tello:2010am,Nemevsek:2011aa,Chakrabortty:2012mh,Dev:2013vxa}.
Detailed studies which take into account potential signals with $\sqrt{s} = 14$ TeV at the LHC
conclude that heavy gauge bosons and neutrinos can be found with up to 4 and 1 TeV,
respectively, for typical LR scenarios \cite{Nemevsek:2011hz,Ferrari:2000sp}. Such a relatively low (TeV) scale of
the heavy sector is theoretically possible, even if GUT gauge unification is demanded, for
a discussion, see e.g. \cite{Shaban:1992he,Chakrabortty:2009xm}.

In this paper  we consider Left-Right symmetric model based on the $SU(2)_L \otimes SU(2)_R \otimes U(1)_{B-L}$ gauge
group \cite{Mohapatra:1974gc} in its most restricted form, so-called Minimal Left-Right Symmetric
Model (MLRSM).  We choose to explore  the most popular version of the model
with Higgs representations -- a bi-doublet $\Phi$ and two (left and right) triplets $\Delta_{L,R}$ \cite{Gunion:1989in,Duka:1999uc}.  We also assume that the
vacuum expectation value of the left-handed triplet $\Delta_{L}$ vanishes, $\langle\Delta_{L}\rangle =0$ and the CP symmetry can be violated  by
complex phases in the quark and lepton mixing matrices. Left and right
gauge couplings are chosen to be equal, $g_L=g_R$. 
For reasons discussed in \cite{Czakon:2002wm} and more extensively in \cite{Gluza:2002vs}, we discuss see-saw  diagonal light-heavy neutrino mixings. 
It means that $W_1$ couples mainly to light neutrinos, while $W_2$ couples to the heavy ones.  
$Z_1$ and $Z_2$ turn out to couple to both of them \cite{Gluza:1993gf,Duka:1999uc}. $W_L-W_R$ mixing is allowed and is very small, 
$\xi \leq 0.05$  \cite{Beringer:1900zz}, the most stringent data comes from astrophysics through the supernova explosion analysis \cite{Mohapatra:1988tm}.
In our last paper we considered low energy constraints on such a model assuming $\kappa_2=0$, i.e., $\xi=0$ \cite{Chakrabortty:2012pp}, we do the same here.
Moreover, in MLRSM  $\tan{2\xi}=-\frac{2\kappa_1 \kappa_2}{v_R^2}$, which is really negligible for $v_R \geq 5$ TeV, as dictated by Eq.~(\ref{mw2lhc}), 
where $\kappa_1,\kappa_2(v_R)$ are the vacuum expectation values of $\Phi(\Delta_R)$.

We think that it is worth to show how the situation looks like if we stick to the popular and to a large extent conservative version of the model (MLRSM), giving  
candle-like benchmark numbers for possible signals at the LHC.  We should also be aware of the fact, that there are relations between model parameters 
in the Higgs, gauge and neutrino sectors \cite{Czakon:1999ue,Czakon:1999ga,Duka:1999uc,Chakrabortty:2012pp} and it needs further detailed studies.
For estimation and discussion of observables which are able to measure final signals in the most efficient way, calculation of dominant tree level signals is sufficient at the moment.
Production processes are calculated and relevant diagrams are singled out using CalcHEP \cite{Belyaev:2012qa}. 
For general analysis, multi lepton codes  ALPGEN \cite{Mangano:2002ea}, PYTHIA \cite{Sjostrand:2006za},
Madgraph \cite{Alwall:2011uj} are used. Feynman rules are generated with our version of 
the package using FeynRules \cite{Christensen:2008py,Degrande:2011ua}.  
The backgrounds for multi lepton signals (3 and 4 leptons) are estimated  using ALPGEN-PYTHIA. 

In this paper we have grabbed the impact of the relatively light charged scalars in the phenomenology of Left-Right symmetric model. We first discuss how the decay branching ratios of $W_2,\;Z_2,$ and $N_R$ are affected by the presence of these light charged scalars. Then we note down the possible interesting processes within MLRSM. We study the production and decay modes of the charged scalars. 
We have provided some benchmark points where we have performed our simulations to make a realistic estimation of the signal events over the SM backgrounds. Our study is based on the reconstruction of the invariant masses of the final state leptons and their mutual separations from where we have shown how we can track the presence of doubly charged scalars. We also note down the impact of the charged scalars in the Higgs to di-photon decay rates. Then we conclude and give an outlook.
    
\section{MLRSM processes with charged Higgs boson particles at the LHC}

There are already severe limits on the heavy gauge boson masses, 
Eq.~(\ref{mw2lhc}), which infer that scale in which the right $SU(2)$ gauge sector is broken at $v_R>5$ TeV (for approximate relations between gauge boson masses and $v_R$, see for example Eq.~(2.4) in \cite{Chakrabortty:2012pp}).
This is already an interesting situation as for such heavy gauge bosons most of the effects connected 
with them decouple in physical processes at collider physics.
Then there is a potential room to go deeper and estimate more sensitive  Higgs boson contributions. Of course, the effects coming from 
the scalar sector depend crucially also on their masses. Smaller the Higgs boson masses, larger effects are expected.
The question is then: how small their masses can be by keeping the right scale $v_R$ large? In the paper we assume light charged scalar masses up to 600 GeV, this choice of masses will be justified when production cross sections are considered.

\begin{figure}[h]
\begin{center}
\includegraphics[width=7.5cm , angle=0]{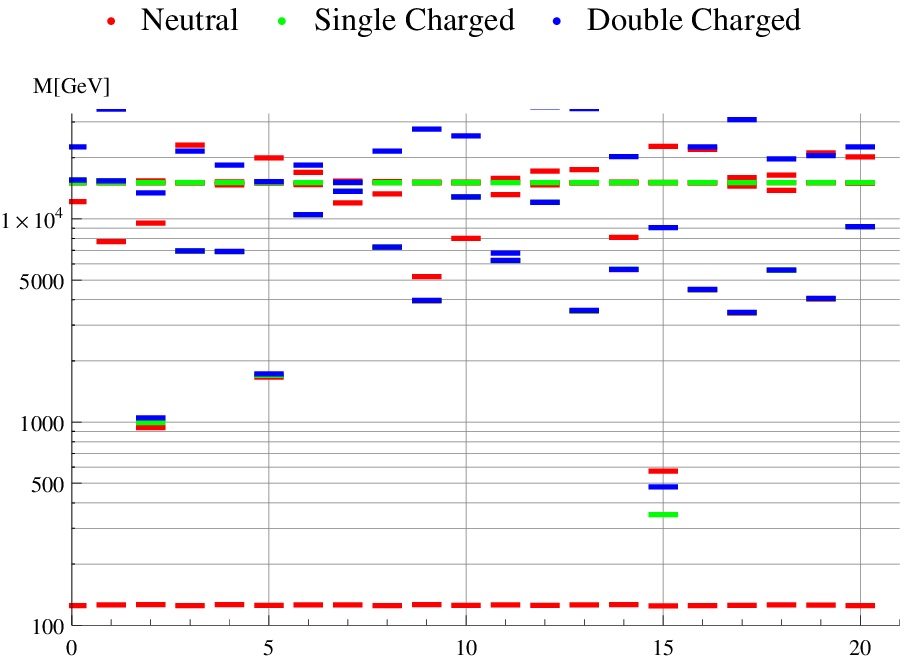} 
\includegraphics[width=7.5cm , angle=0]{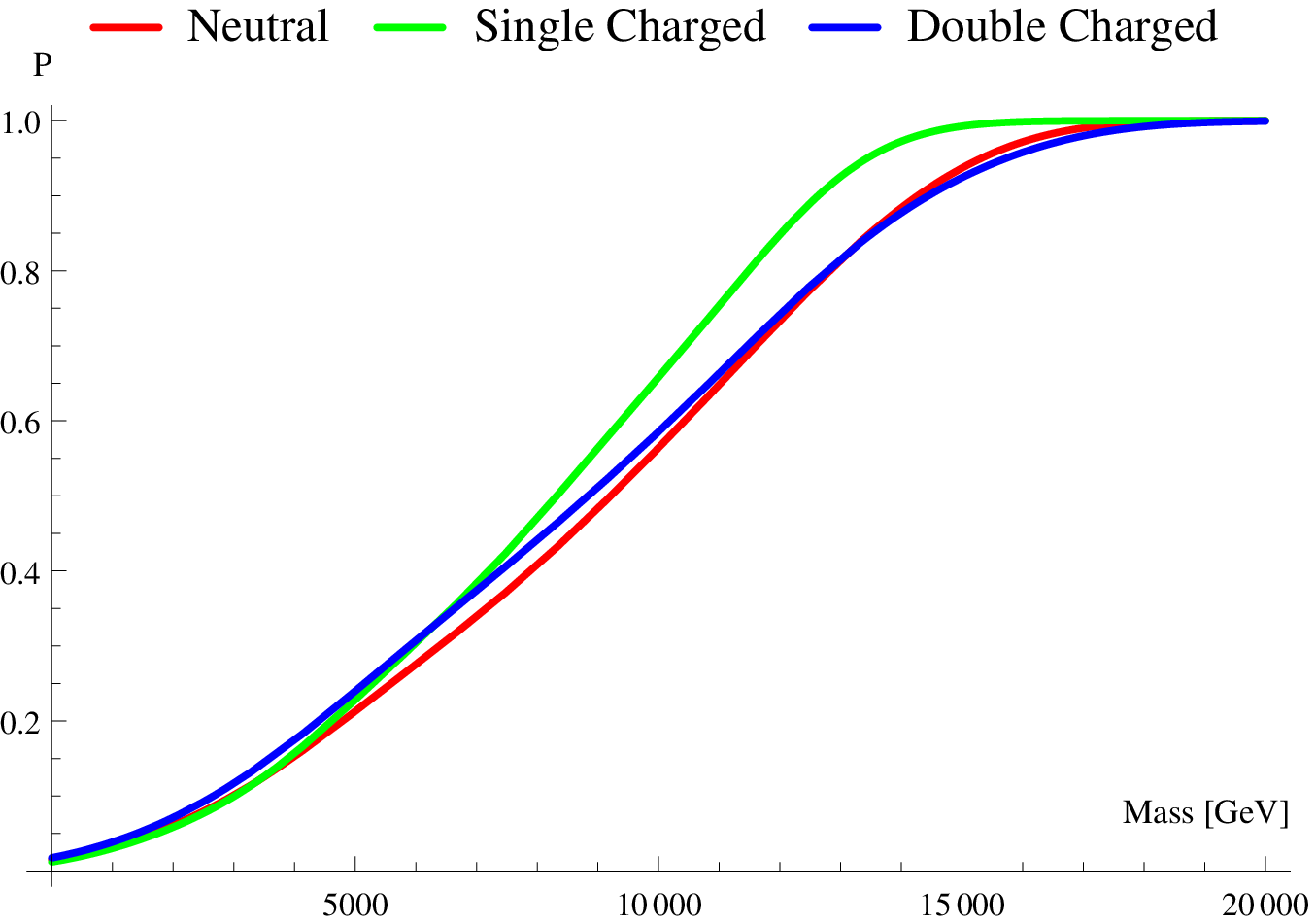} 
\caption{{\underline{On left:}} an example of 20 Higgs mass spectra  obtained by randomly chosen Higgs potential parameters. The constrain on the lowest neutral Higgs mass Eq.(\ref{HiggsH0}) was imposed and the bounds coming from FCNC were taken into account. {\underline{On right:}} cumulative distribution function $P$ of the lowest mass of singly and doubly charged and next to lightest neutral scalars. For both figures, $v_R=8$ TeV. 
}
\label{higgsspec}
\end{center}
\end{figure}

  The point is that all Higgs scalars are naturally of the order of $v_R$, in addition, neutral Higgs boson scalars $A_1^0$ and $H_1^0$ contribute to FCNC effects (see the Appendix) and must be large, above 10 TeV (see however \cite{Guadagnoli:2010sd} for alternative solutions).   
  Let us see then if theoretically charged Higgs bosons can have masses below 1 TeV. 
In the model which we consider in this paper we assume that the Higgs potential 
is given as in \cite{Gunion:1989in,Duka:1999uc}, we will also use the same notation, for details on the parametrisation of the Higgs scalar mass spectrum, see the Appendix.
This model includes a number of parameters:  
$\mu_1, \mu_2, \mu_3, \rho_1, \rho_2, \rho_3, \rho_4, \alpha_1, \alpha_2, \alpha
_3$, $\alpha_4$, $\lambda_1,\lambda_2,\lambda_3,\lambda_4$.
 The exact Higgs mass spectrum is calculated numerically. Minimisation conditions are used to get values of dimensionful mass parameters 
 $\mu_1$, $\mu_2$ and $\mu_3$ which can be arbitrarily  large, all other parameters are considered as free, but limited to the perturbative bound\footnote{Which is equal to $4 \pi$, otherwise   proper analysis of the Higgs potential with radiative corrections to determine perturbative regions would be needed.}, 
 $|\rho_i|,|\alpha_i|, |\lambda_i| <10$. 
  It is  assumed that the lightest neutral Higgs particle is the boson  discovered by ATLAS and CMS collaborations. We have taken its mass to lie in the range 
 \begin{equation}\label{HiggsH0}
 124.7\;{\mbox{GeV}} < M_{H_0^0} < 126.2\;{\mbox{GeV}}.
 \end{equation}
 \\
 An example set of generated mass spectra of Higgs bosons for $v_R=8$ TeV is presented in Fig.~\ref{higgsspec} (left figure). 
Mass spectra have been obtained by varying uniformly the Higgs potential parameters in a range (-10,10). We have also taken into account the bounds on neutral Higgs bosons obtained from FCNC constrains assuming  
$m_{A_1^0},m_{H_1^0}>15$ TeV by fixing $\alpha_3=7.1$ (see Appendix A).
The spectra which did not fulfill  relation (\ref{HiggsH0}) were rejected. 
Altogether we have  6 neutral, 2 singly charged and 2 doubly charged Higgs boson particles in the MLRSM. 
The figure includes possible spectra of singly and doubly charged as well as neutral Higgs bosons.  Some of them can be degenerated or nearly degenerated.

This study shows that although the Higgs particles naturally tend to have masses of the order of the $v_R$ scale, it is still possible to choose 
the potential parameters such that some of the scalar particles can have masses much below 1 TeV (spectrum 15).
To discuss spectra more quantitatively, the  cumulative distribution function $P$ of the lowest masses of singly and doubly charged  
and next to lightest  neutral scalar particles are plotted on right Fig.~\ref{higgsspec}, again for the same conditions as before and $v_R=8$ TeV.
  These results show that for $v_R=8$ TeV a fraction of 
the parameter space that gives lightest scalar masses below 1 TeV is at the level of 4\%.   
It means that it is possible to generate the low mass spectra of Higgs boson masses in MLRSM keeping large $v_R$ scale. However, what can not be seen on those plots is that in MLRSM not all four charged Higgs bosons can simultaneously be light. It is a case for $H_1^\pm$, $H_1^{\pm \pm}$ and 
$H_2^{\pm \pm}$, for details, see the Appendix. The remaining charged scalar $H_2^\pm$ is of the order of the $v_R$ scale, so its effects at LHC is negligible, to make it lighter would require to go beyond MLRSM.  
For a book keeping, we keep this particle in further discussion. 
If its mass at some points is assumed to be small (so we go beyond MLRSM), we denote it with a tilde, $\tilde{H}_2^\pm$.  Its coupling is kept all the time as in MLRSM (why it can be so is discussed shortly in the Appendix). 

In this paper we consider only the processes where charged Higgs particles can be produced directly as shown in the 
Table~\ref{tabrange1}, first column.

\begin{table}[h!]
\begin{tabular}{|c|c|c|}
\hline \hline
Primary production & Secondary production & Signal  \\
  \hline  
 I. $~~~ H_1^{+} H_1^{-}$  & $\ell^{+} \ell^{-} \nu_L \nu_L$ & 
 $\ell^{+} \ell^{-} \oplus MET$  
 \\
 \hline
 -- & $\ell^{+} \ell^{-} N_R N_R$ &  
 depends on $N_R$ decay modes  \\
 \hline
 -- & $\ell^{+} \ell^{-} \nu_L N_R$ & 
 depends on $N_R$ decay modes \\
\hline
 II. $~~~ H_2^{+} H_2^{-}$  & $\ell^{+} \ell^{-} \nu_L \nu_L$ & $\ell^{+} \ell^{-} \oplus MET$  \\
 \hline
 -- & $\ell^{+} \ell^{-} N_R N_R$ & 
 depends on $N_R$ decay modes  \\
 \hline
 -- & $\ell^{+} \ell^{-} \nu_L N_R$ &   
 depends on $N_R$ decay modes  \\
\hline
III. $~~~ H_1^{++} H_1^{--}$  & -- &  $\ell^{+} \ell^{+} \ell^{-} \ell^{-}$  \\
 \hline
 -- &  $H_1^+ H_1^+ H_1^- H_1^-$  & See I \\
 \hline 
  -- &  $H_1^{\pm} H_1^{\pm} H_2^{\mp} H_2^{\mp}$  & See I \& II  \\
 \hline
 -- &  $H_2^+ H_2^+ H_2^- H_2^-$  & See II  \\  
 \hline
  -- & $W_i^+ W_i^+ W_j^- W_j^-$ & depends on $W$'s decay modes   \\
   \hline
 IV. $~~~ H_2^{++} H_2^{--}$  & -- &  $\ell^{+} \ell^{+} \ell^{-} \ell^{-}$  \\
\hline
 -- &  $H_2^+ H_2^+ H_2^- H_2^-$  & See II  \\
 \hline 
  -- &  $H_1^{\pm} H_1^{\pm} H_2^{\mp} H_2^{\mp}$  & See I \& II   \\
 \hline
 -- &  $H_1^+ H_1^+ H_1^- H_1^-$  & See I \\
 \hline 
 --  & $W_i^+ W_i^+ W_j^- W_j^-$ &  depends on $W$'s decay modes  \\
\hline
V. $~~~H_1^{\pm \pm} H_1^\mp$ &  -- & $\ell^\pm \ell^\pm \ell^\mp \nu_L$ \\
\hline
VI. $~~~H_2^{\pm \pm} H_2^\mp$ &  -- & $\ell^\pm \ell^\pm \ell^\mp \nu_L$  \\
\hline
VII. $~~~H_1^{\pm} Z_i,H_1^{\pm} W_i$ &  -- & See I \& $Z_i, W_i$ decay modes   \\
\hline
VIII. $~~~H_2^{\pm} Z_i, H_2^{\pm} W_i$ &  -- & See II \&  $Z_i, W_i$ decay modes   \\
\hline
IX. $~~~H_1^{\pm} \gamma$ &  -- & See I   \\
\hline
X. $~~~H_2^{\pm} \gamma$ &  -- & See II  \\
\hline
\hline
\end{tabular}
\caption{ Phenomenologically interesting MLRSM processes at the LHC with primarily produced charged scalar particles and possible final signals.
Here $\gamma$ denotes a photon. $\nu_L=\nu_1,\nu_2,\nu_3$ are SM-like light massive neutrino states and $N_R=N_{4,5,6}$ are heavy neutrino massive 
states dominated by right-handed weak neutrinos. From now on we will denote $N_R \equiv N$. Here $\ell$ represents light charged leptons $e,\mu$.}
\label{tabrange1}
\end{table}

\begin{figure}[t]
\begin{center}
\includegraphics[width=7.5cm , angle=0]{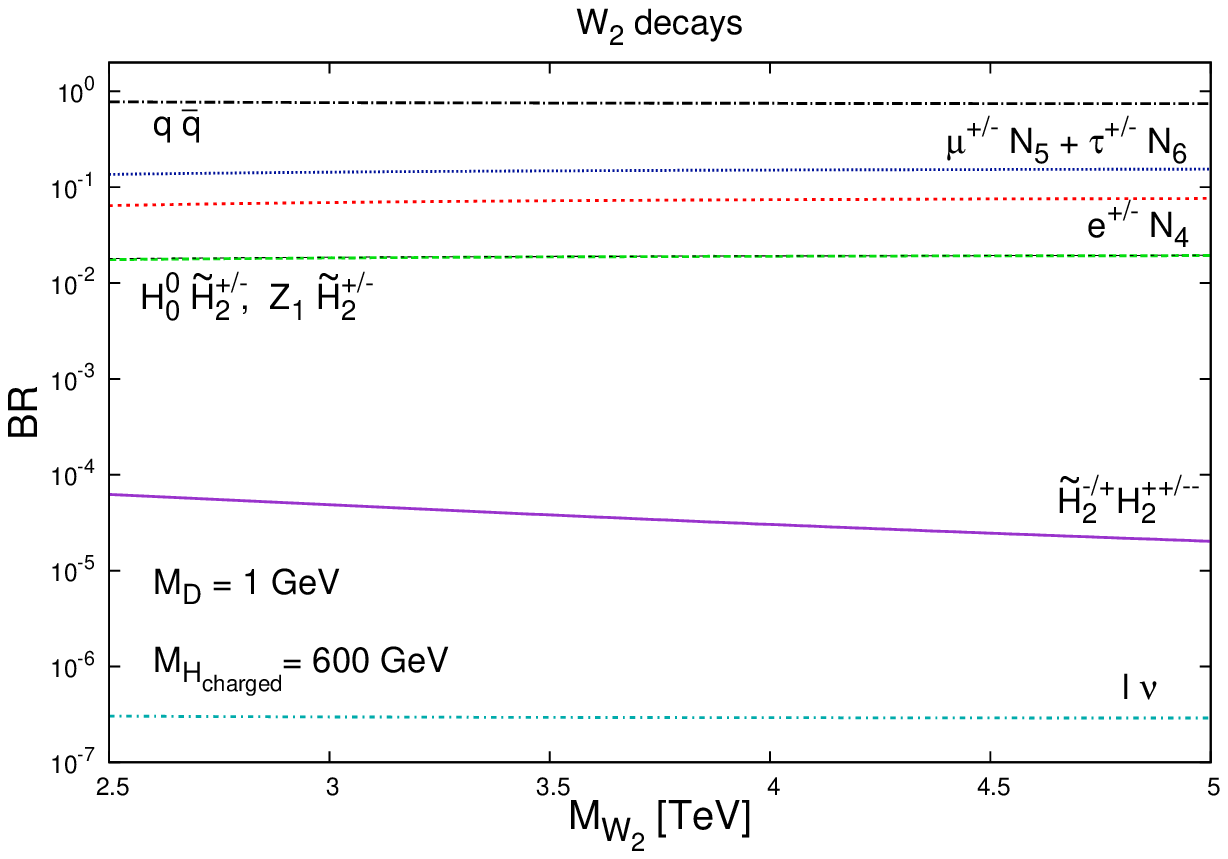} 
\includegraphics[width=7.5cm , angle=0]{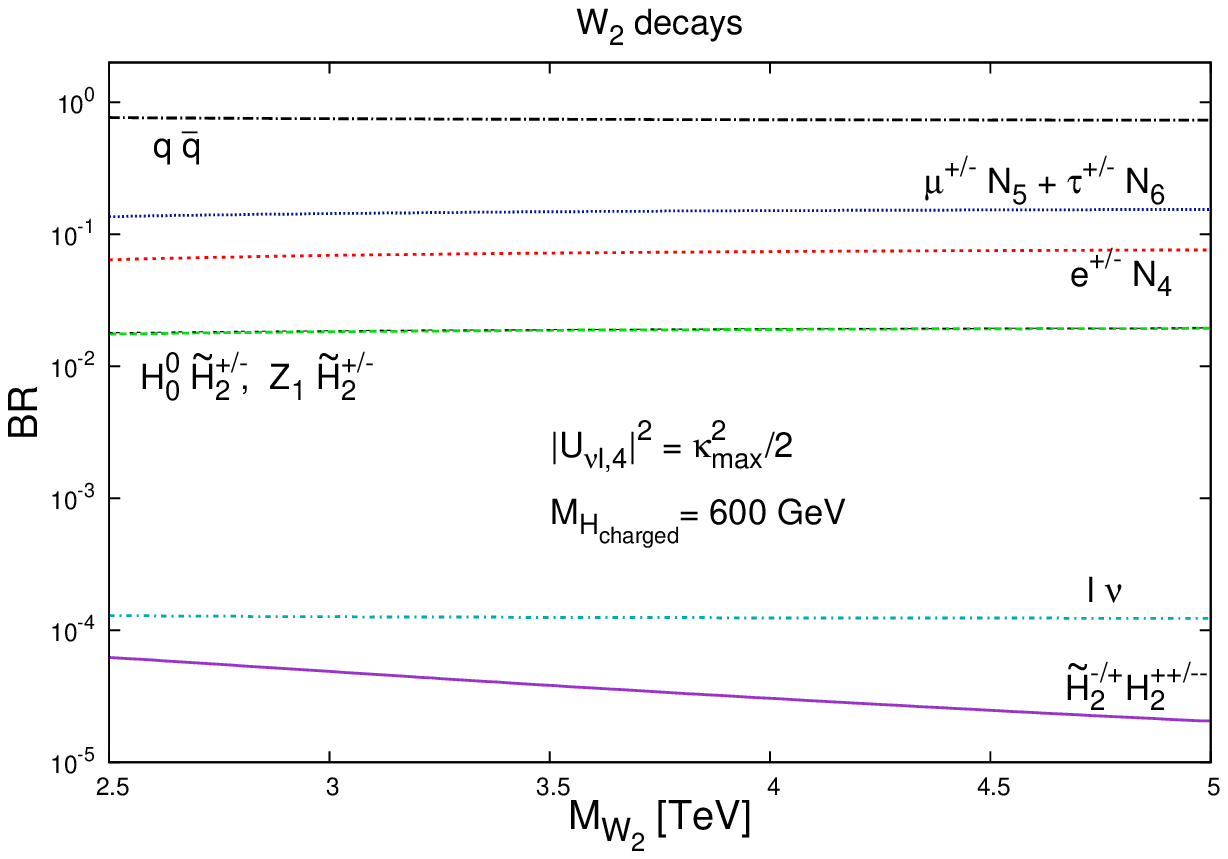} 
\caption{Branching ratio for $W_2$ decay with relatively light charged scalars. Here we put $M_{N_4}=M_{N_5}=1$ TeV, $M_{N_6}=800$
GeV. Symbol $q \bar{q}$ on this and next plots stands for a sum of all quark flavours,
$q \bar{q}\equiv\sum\limits_{i, i'=u,d,s,b,c,t} q_i \bar{q_{i'}}$. Similarly,
$l \nu \equiv \sum\limits_{i=1}^3 l_{i}\nu_i$.}
\label{brmw2}
\end{center}
\end{figure}

\begin{figure}[h]
\begin{center}
\includegraphics[width=7.5cm , angle=0]{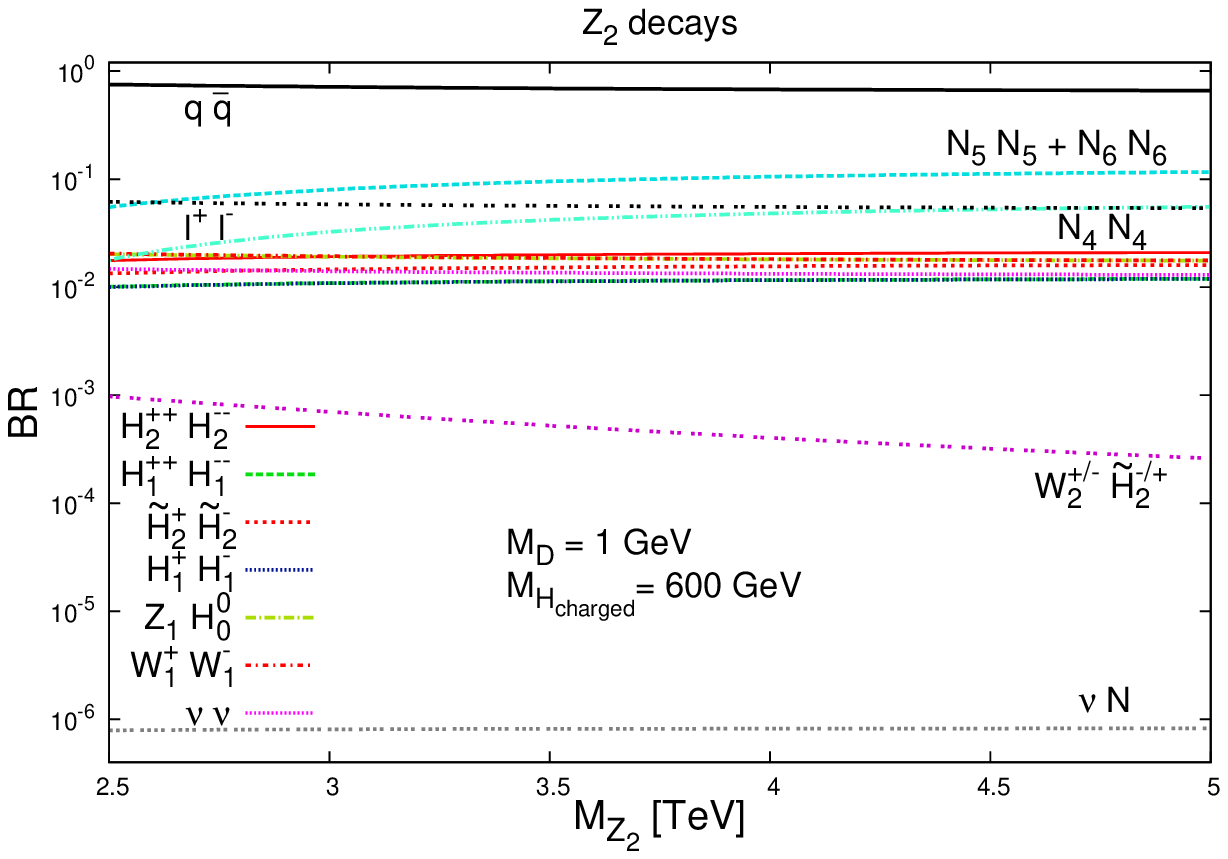} 
\includegraphics[width=7.5cm , angle=0]{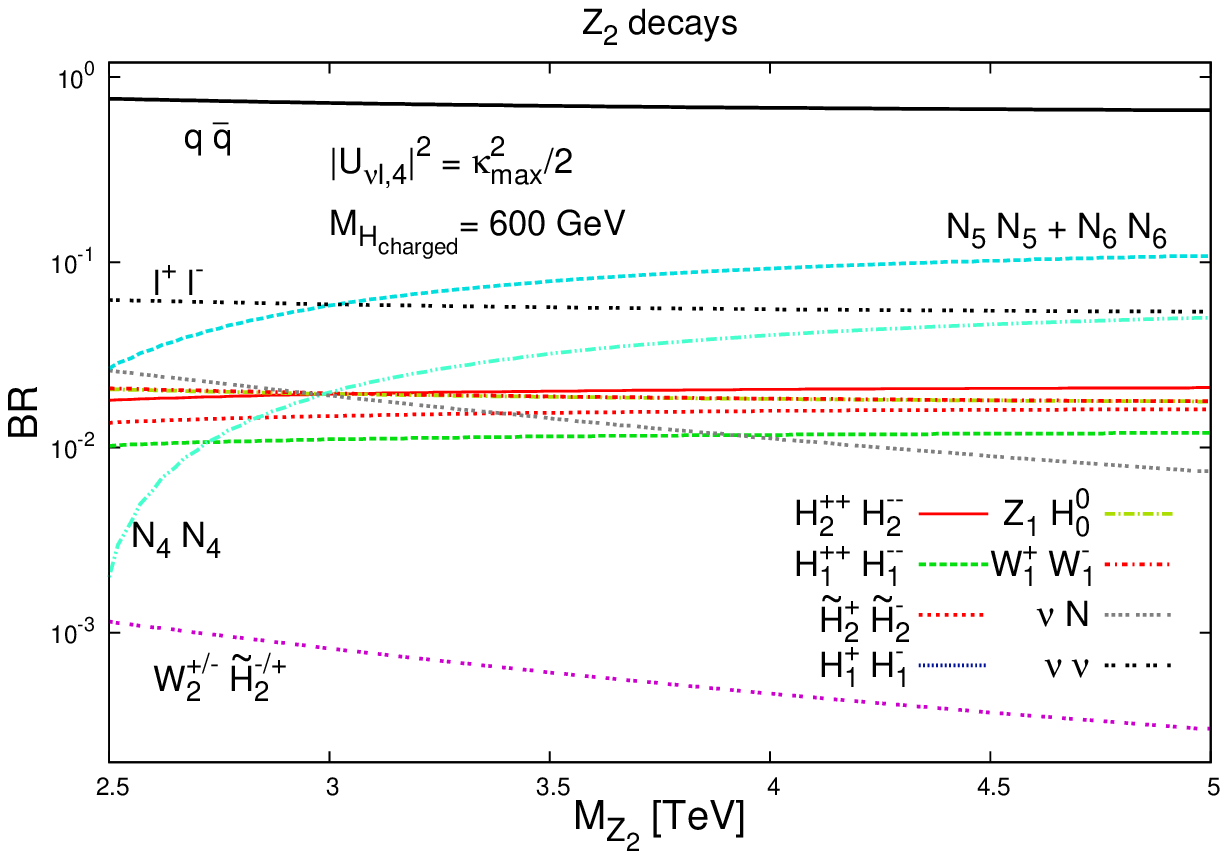} 
 
\caption{Branching ratio for $Z_2$ decay with relatively light charged scalars. Here $\nu \nu \equiv \sum\limits_{i=1}^3 \nu_i \nu_i$ and 
$\nu N \equiv \sum\limits_{i=1}^3 \nu_{i} N_{i+3}$.}
\label{brmz2}
\end{center}
\end{figure}

\begin{figure}[h]
\begin{center}
\includegraphics[width=7.5cm, angle=0]{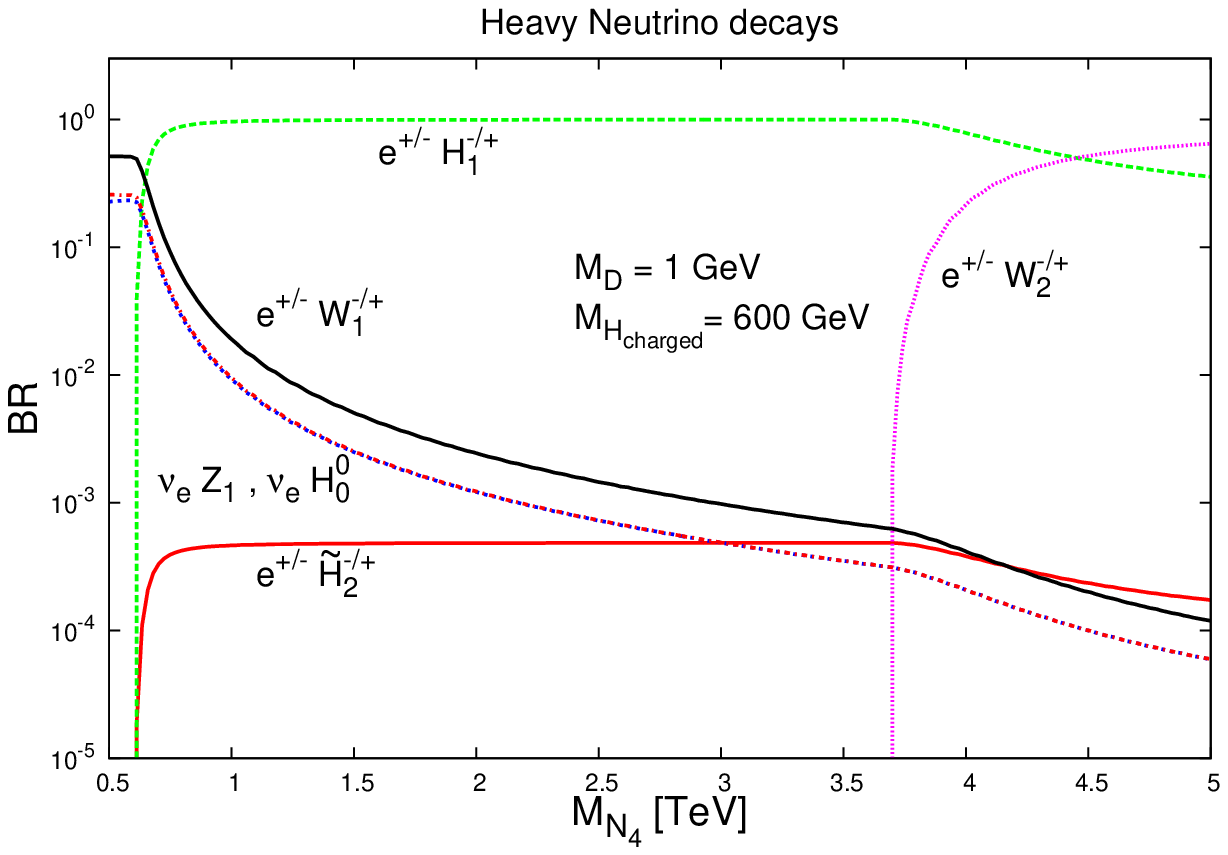}
\includegraphics[width=7.5cm, angle=0]{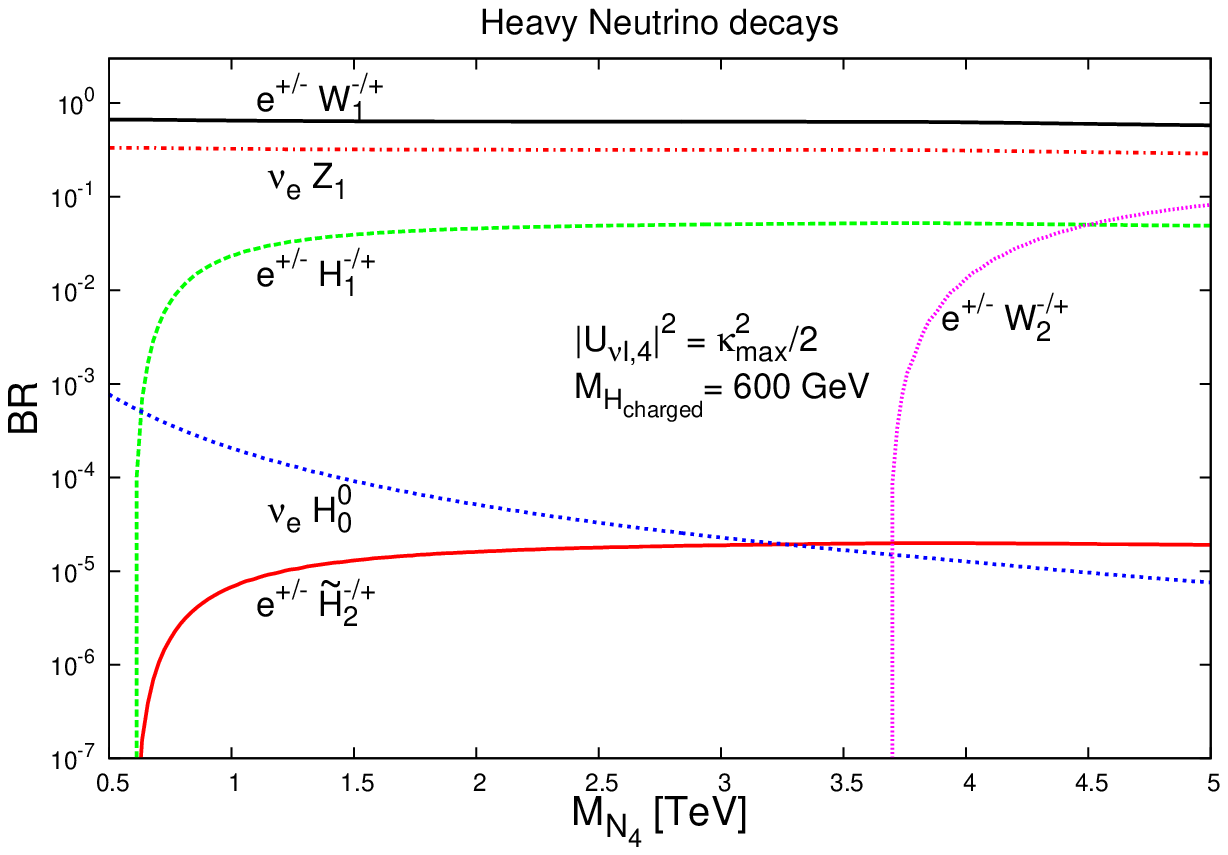} 
\caption{Branching ratios for $N_4$ decay with relatively light charged scalars.}
\label{brmn}
\end{center}
\end{figure}

The decay branching ratios for heavy neutrino states $N$ and heavy gauge bosons $(W_2,Z_2)$ in MLRSM which determine both secondary production and final signals in the last column of this table are given in \cite{Chakrabortty:2012pp}. However, with assumed light charged Higgs particles, new decay modes are potentially open, and discussion must be repeated. 
Results are given in Figs.~\ref{brmw2}, \ref{brmz2}, and \ref{brmn}.
As can be seen from Fig.~\ref{brmw2}, contribution of charged scalars to the total decay width of $W_2$ is at the percent level. 
Here more important are heavy neutrino decay modes\footnote{Some processes in the Table~\ref{tabrange1} depend strongly on the light-heavy (LH) neutrino mixing scenarios.}. Different scenarios for LH neutrino mixings 
\cite{Chakrabortty:2012pp} are discussed, i.e., see-saw mechanisms where $|U_{\nu_i j}| \simeq \frac{|\langle M_D \rangle|}{M_{N_j}}  \delta_{i,j-3},\; i=1,2,3,\;j=4,5,6$ and scenarios where LH neutrino mixings are independent of neutrino masses: 
$\sum_{j=4,5,6} U_{\nu_1, {j-3}} U_{ \nu_1, j-3}^{\ast} = U_{\nu_1, 4} U_{\nu_1, 4}^{\ast} \leq 0.003 \equiv \kappa^2_{max}$ \cite{delAguila:2008pw}.  In a case of many heavy neutrino states (as in MLRSM), taking into account constraints coming from neutrinoless double-beta decay experiment, this limit becomes $\kappa^2_{max}/2$  
\cite{Gluza:1997ts,Gluza:1995ix,Gluza:1995ky}. For $W_2$ decays different LH neutrino mixing scenarios affect only light neutrino $\nu l$ channel for which BR is small, anyway. 

For the $Z_2$ decays, 
Fig.~\ref{brmz2}, four channels with charged Higgs bosons, namely $H_1^{++}  H_1^{--}$, $H_1^+ H_1^-$, $H_2^{++}  H_2^{--}$, and $\tilde{H}_2^+ \tilde{H}_2^-$, contribute to the decay rate in a percentage level. The quark decay modes dominate, and the second important are the heavy neutrino decay modes.

The most interesting situation is for the decays of heavy neutrinos. Here $H_1^+$ decay mode is the largest in see-saw scenarios. 
The reason is that in case of Yukawa coupling, say  $H_1^+-N-e$, the change in LH neutrino mixing is compensated by the proportionality of the coupling to the heavy neutrino mass, which is not the case for the gauge $N-e-W$ and $N-\nu-Z$ couplings. That is why $e W$ and $\nu Z$ decay modes are relevant only in scenarios where LH neutrino mixings are independent of the heavy neutrino masses and are close to the present experimental limits. Large charged Higgs boson decay mode of the heavy neutrino can influence the ``golden" $pp \to e N$ process \cite{Keung:1983uu,Ferrari:2000sp,Nemevsek:2011hz,Das:2012ii,Chen:2013foz,Chakrabortty:2012pp}.
 
For typical see-saw cases when charged Higgs boson masses are very large, standard model modes dominate: $N \to
e W_1$ and $N \to \nu_L Z_1$ if $M_{N} <M_{W_2}$ whereas $N \to e W_2$ if 
$M_{N} > M_{W_2}$. In scenarios with large LH neutrino mixings the standard modes dominates independently of the heavy neutrino and $W_2$ masses\footnote{Relevance of see-saw LH mixings at the LHC has been discussed lately in \cite{Chen:2013foz}.}.
Finally, let us note that in typical Type I see-saw scenarios the TeV scale of heavy neutrino masses implies $m_D \sim 10^{-6}$ GeV to accomplish light neutrino masses at the eV level. In this situation nothing happens to the left plots in Figs.~\ref{brmw2}, \ref{brmz2}, and \ref{brmn} apart from the fact that $l \nu$, $l N$ and $\nu Z$ channels will disappear completely there.

In the case of heavy gauge boson decays, quarks dominate and jets will be produced while for SM-like gauge bosons hadronic decay branching is around 70\%.
That is why typical final signals for reactions I and II in Table~\ref{tabrange1} are two or four jets plus missing energy. 
There are only two cases without missing energy:
\begin{equation}
H_{1(2)}^+ H_{1(2)}^- \to \ell^+ \ell^- N N \to \ell^+ \ell^- W_{m}^{\pm} \ell^{\mp} W_{n}^{\pm} \ell^{\mp} \to jj jj \ell^+ \ell^- \ell^{\mp}\ell^{\mp},
\label{h12h12a}
\end{equation}
and
\begin{equation}
H_{1(2)}^+ H_{1(2)}^- \to \ell^+ \ell^- N N \to \ell^+ \ell^-   W_{m}^{\pm} \ell^{\mp} W_{n}^{\mp} \ell^{\pm} \to jj jj \ell^+ \ell^- \ell^{\pm}\ell^{\mp}.
\label{h12h12b}
\end{equation}
 
However, as we can see from the table, the cleanest signals are connected with doubly charged Higgs particles, that is why we focus on them in this paper. For some related discussions on doubly charged scalars, see e.g. \cite{Huitu:1996su,PhysRevD.40.1521,Rizzo:1981xx,Maalampi:2002vx,delAguila:2008cj,Melfo:2011nx,delAguila:2013yaa,Babu-Ayon,delAguila:2013mia}. The processes  Eqs.~(\ref{h12h12a}) and (\ref{h12h12b}) with four charged leptons plus jets will be considered elsewhere.

For processes III-X important are charged Higgs boson decay modes. 
For doubly charged Higgs particles possible decay modes are

\begin{equation}
\begin{array}{ll}
{\rm (i)} &H_1^{\pm \pm} \rightarrow l^{\pm} l^{\pm}, \; \\
{\rm (ii)} &H_1^{\pm \pm} \rightarrow H_1^{\pm} W_1^{\pm}; \\
{\rm (iii)} &H_2^{\pm \pm} \rightarrow l^{\pm} l^{\pm},\; \\
{\rm (iv)} &H_2^{\pm \pm} \rightarrow H_2^{\pm} W_2^{\pm}; \\
{\rm (v)} &H_2^{\pm \pm} \rightarrow W_2^{\pm} W_2^{\pm}; \\
{\rm (vi)} &H_2^{\pm \pm} \rightarrow H_2^{\pm} W_1^{\pm}; \\
\end{array}
\label{brll}
\end{equation}
 where $l = e, \mu, \tau$.
 
Apart from the above decay modes,  the other possibilities for the doubly charged scalars can be  
\begin{equation}
\begin{array}{ll}
{\rm (vii)} &H_2^{\pm \pm} \rightarrow H_1^{\pm} H_1^{\pm}, \; \\
{\rm (viii)} &H_2^{\pm \pm} \rightarrow H_2^{\pm} H_2^{\pm}; \\
\end{array}
\end{equation}
when they are not degenerate with the  singly charged ones. But for nearly or exact degenerate case, the charged scalars dominantly decay through leptonic modes and here kinematics play a role too.

\begin{figure}[h]
\begin{center}
\includegraphics[width=6cm , angle=270]{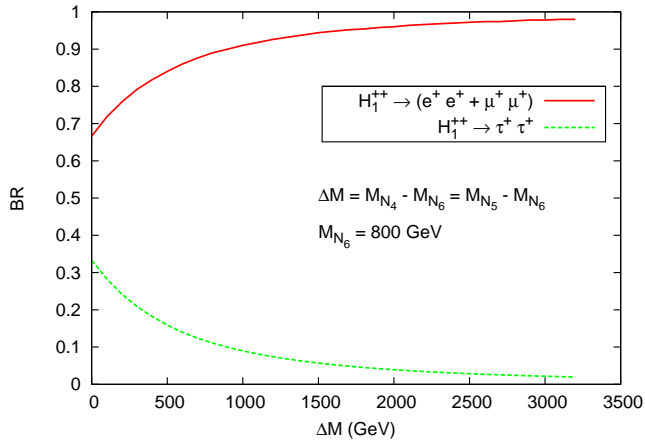}
\end{center} 
\caption{Branching ratios for the decay modes $(e^+ e^+ + \mu^+ \mu^+)$ and $ \tau^+ \tau^+$ of the doubly charged scalars as a function of $\Delta M$, where
$\Delta M = M_{{N_4}} - M_{{N_6}} = M_{{N_5}} - M_{{N_6}}$. We have kept fixed $M_{{N_6}} = 800$ GeV. Note that the BRs of both the doubly charged scalars ($H_1^{++}$ and $H_2^{++}$) 
are the same in scenarios where $M_{W_2}>> M_{H^{\pm \pm}}$ and $M_{H^{\pm \pm}} \simeq M_{H^{\pm}}$.   }
\label{bree}
\end{figure}
 
 Fig.~\ref{bree} shows a scenario in which pure leptonic decay modes can be realised. The crucial factor is the Yukawa coupling which depends (indirectly) on heavy right-handed neutrino mass.
If heavy neutrino masses are degenerate then democratic scenario is understood where all leptonic channels are the same (i.e. ${\rm BR}(H^{\pm \pm} \to e^\pm e^\pm)\simeq$ 33\%).

Typically, as can be seen from Fig.~\ref{bree}, for right-handed neutrino masses to be  1 TeV, 1 TeV and 800 GeV for $N_4,N_5,N_6$ respectively, the branching ratios are the following  \\
\begin{equation}
\begin{array}{l}
{\rm BR} (H_{1/2}^{\pm \pm} \rightarrow e^{\pm} e^{\pm}) = 37.9\%,\\
{\rm BR} (H_{1/2}^{\pm \pm} \rightarrow \mu^{\pm} \mu^{\pm}) = 37.9\%,\\
{\rm BR} (H_{1/2}^{\pm \pm} \rightarrow \tau^{\pm} \tau^{\pm}) = 24.2\%.
\end{array}
\label{brll}
\end{equation}

If the first two generations neutrinos ($N_4,N_5$) have masses above $\sim 4$ TeV, $\tau$ decay mode is practically irrelevant. From the discussion it is also clear, that \emph{one of the decay modes can dominate if only one of the right-handed neutrino masses is much bigger than remaining two heavy neutrino states}. Limits in Eq.~(\ref{mhpplimit})
assume 100\% leptonic decays, in our case, taking into account Fig.~\ref{bree}, Eq.~(\ref{brll}) and results given in \cite{CMS:2012kua,ATLAS:2012hi}, mass limits are much weaker, at about 300 GeV, see e.g. Fig.~3 in \cite{ATLAS:2012hi}.

 For decays of singly charged $H_1^\pm$ scalars situation is analogical as for doubly charged scalars (possible decay modes to neutral $H_1^0$ and $A_1^0$ scalars are negligible for $M_{H_1^0},M_{A_1^0}>> M_{H_1^\pm}$, as dictated by FCNC constraints).

 $\tilde{H}_2^{\pm}$ decays hadronicaly, namely, for $100\;{\rm GeV}<M_{\tilde{H}_2^{\pm}}<200$ GeV
\begin{equation}
\begin{array}{l}
{\rm BR}(\tilde{H}_2^{+} \rightarrow c \bar{s} ) = 95\%,\\
{\rm BR} (\tilde{H}_2^{+} \rightarrow c \bar{d} ) = 5 \%,
\end{array}
\end{equation}
and ${\rm BR} (\tilde{H}_2^{+} \rightarrow t \bar{b} ) \sim 100 \%$  for $M_{\tilde{H}_2^{\pm}}> 200$ GeV.

\subsection{Primary production of heavy charged Higgs bosons at the LHC}

Below different processes involving solely charged scalar productions are classified.
In analysis which follow $v_R = 8000$ GeV to respect with a large excess the present exclusion limits on $W_2^{\pm}$, and $Z_2$ masses. 
SM-Higgs like mass is set to 125 GeV, masses of neutral scalar particles are set at very high limit ($\sim$ 10 TeV). In this way, as already discussed, scenarios are realised with relatively light (hundreds of GeV) charged Higgs bosons while remaining  non-standard  particles within MLRSM are much heavier.  
All cross sections given in this section are without any kinematic cuts, those will be considered with final signals and distributions in section 
\ref{sim}.

\subsubsection{$pp \to H_1^+ H_1^-$ and $pp \to H_2^+ H_2^-$} \label{subsubsection:pptoHmHp}

\begin{figure}[h]
\begin{center}
\includegraphics[width=7cm, angle=-90]{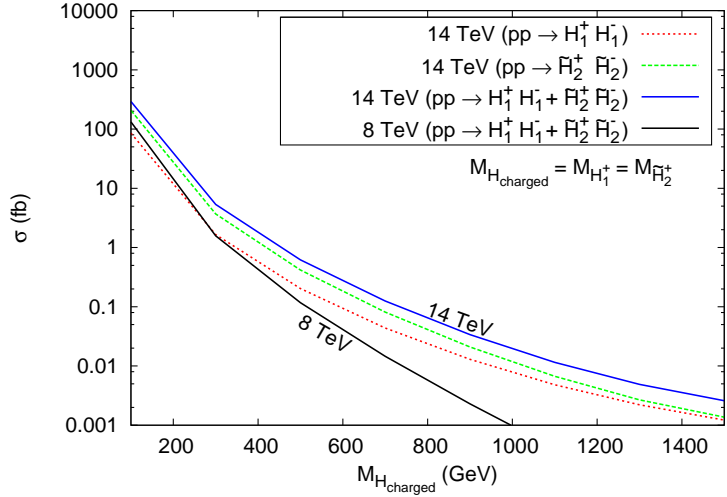}
\caption{Production cross sections  for $pp \to H_1^+ H_1^-$ and $pp \to \tilde{H}_2^+ \tilde{H}_2^-$ processes without imposing kinematic cuts.}
\label{fig:prodhphm}
\end{center}
\end{figure}

The cross section for singly charged scalar pair production as a function of their mass is given in Fig.~\ref{fig:prodhphm}. This process is dominated by s-channel $\gamma,Z_1$ and t-channel quark exchange diagrams.  
Contributions coming from s-channel $H_0^0,Z_2$ and $H_1^0$ bosons are negligible   for considered MLRSM parameters. 
For singly charged scalar mass equals to 400 GeV, the cross sections are 
(as discussed in Section 2, $H_2^\pm$ Higgs boson is assumed to be light and we denote it here with a tilde, for $M_{H_2^\pm}>> 1$ TeV the considered cross section is negligible, $\sigma(pp\rightarrow \tilde{H}^{\pm}_{2}\tilde{H}^{\mp}_{2}) \simeq 0$)
\begin{eqnarray} 
\sigma(pp\rightarrow H^{\pm}_{1}H^{\mp}_{1}) &=& 0.12 (0.52)\; fb, \\
\sigma(pp\rightarrow \tilde{H}^{\pm}_{2}\tilde{H}^{\mp}_{2}) &=& 0.27 (1.12)\; fb, 
\end{eqnarray}
while for singly charged scalar mass equals to 600 GeV are
\begin{eqnarray} 
\sigma(pp\rightarrow H^{\pm}_{1}H^{\mp}_{1}) &=& 0.01 (0.09)\; fb, \\
\sigma(pp\rightarrow \tilde{H}^{\pm}_{2}\tilde{H}^{\mp}_{2}) &=& 0.03 (0.18)\; fb,
\end{eqnarray}
with $\sqrt{s}=8(14)$ TeV. 

Increasing center of mass energy from $\sqrt{s}=8$ TeV to $\sqrt{s}=14$ TeV the cross sections grow by factors $\sim 4 \div 7$, 
depending on masses of charged Higgs bosons. In general cross sections fall down below $0.1~fb$ for masses of charged scalars above approximately 730(420) GeV for $\sqrt{s}=14(8)$ TeV.

\subsubsection{$pp \to H_1^{++} H_1^{--}$ and $pp \to H_2^{++} H_2^{--}$} \label{subsubsection:pptoHmmHpp}

 The dominant contribution to these processes is via neutral s-channel current, i.e., via 
$Z_1$ and $\gamma$. Contributions coming from s-channel $H_0^0,Z_2$ and $H_1^0$ are negligible  for considered MLRSM parameters.

To explore the phenomenological aspects of the doubly charged scalars in the MLRSM model we consider two scenarios. $Scenario~I$ when the doubly charged scalars are degenerated 
in mass, i.e., $M_{H_1^{\pm \pm}} = M_{H_2^{\pm\pm}}$. This scenario is motivated by analysis of the Higgs potential (a detailed study of the Higgs potential and scalar mass spectrum will be presented elsewhere). In $Scenario~ II$ masses are
different, i.e., $M_{H_1^{\pm \pm}} \neq M_{H_2^{\pm\pm}}$. 

 \begin{figure}[htb]
\begin{center}
\includegraphics[width=7cm, angle=-90]{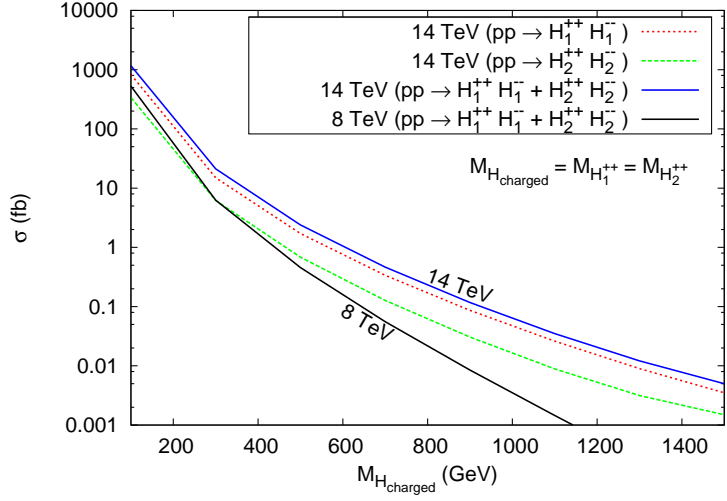}
\caption{Scenario I. Cross sections  for $pp \to H_1^{++} H_1^{--}$ and $pp \to H_2^{++} H_2^{--}$ processes  without imposing kinematic cuts.}
\label{fig:crosssection_pptoHpphmm}
\end{center}
\end{figure}

\subsubsection*{Scenario I, degenerate mass spectrum}
In our analysis we set our benchmark point with both of the doubly charged scalars at the same mass $M_{H_1^{++}} = M_{H_2^{++}} = 400$ GeV.
In this case, the cross section at the LHC without imposing any cut at $ \sqrt{s}=8(14)\;{\rm TeV}$ is 
\begin{equation}\label{eqhpp814}
\sigma(pp\rightarrow (H_{1}^{++} H_{1}^{--}+H_{2}^{++} H_{2}^{--})\rightarrow {\ell_i}^+{\ell_i}^+{\ell_j}^-{\ell_j}^-) = 1.44 (6.06)~fb,
\end{equation}
The contributions to the cross sections from two possible channels are noted for $\sqrt{s}=8(14)\;{\rm TeV}$ as
\begin{eqnarray}
\sigma(pp\rightarrow H_{1}^{++} H_{1}^{--})& = & 1.09 (4.58)~fb, 
\label{eqhpp814a} \\
\sigma(pp\rightarrow H_{2}^{++} H_{2}^{--} )& = & 0.45 (1.86)~fb, 
\label{eqhpp814b}
\end{eqnarray}
where $\ell_{i,j} = e,\mu$.

For $M_{H_1^{++}} = M_{H_2^{++}} = 600$ GeV it is
\begin{equation}
\sigma(pp\rightarrow (H_{1}^{++} H_{1}^{--}+H_{2}^{++} H_{2}^{--})\rightarrow {\ell_i}^+{\ell_i}^+{\ell_j}^-{\ell_j}^-) = 0.14 (0.95)~fb,
\end{equation}
for $\sqrt{s}=8(14)\;{\rm TeV}$.
The contributions to the cross sections from individual channels for $\sqrt{s}=8(14)\;{\rm TeV}$ are as following:
\begin{eqnarray}
\sigma(pp\rightarrow H_{1}^{++} H_{1}^{--} ) &=& 0.11 (0.73)~fb, 
\label{eqhpp814_600}\\
\sigma(pp\rightarrow H_{2}^{++} H_{2}^{--} ) &=& 0.04 (0.28)~fb.
\label{eqhpp814_600_a}
\end{eqnarray}

 The cross sections for pair productions of doubly charged scalars at the LHC with 14 and 8 TeV are given in Fig.~\ref{fig:crosssection_pptoHpphmm}.
From the figure we can see that cross sections fall very rapidly as the masses of the doubly charged scalars increase.
Also the production cross section for $H_1^{\pm \pm}$ is much larger than that for $H_2^{\pm \pm}$ as shown in the figure.
The cross section at $\sqrt{s}=14(8)$ TeV for scalar masses above 920(640) GeV  is $\leq$ 0.1 $fb$. 

\begin{figure}[htb]
\hspace{-0.4cm}
\includegraphics[height=10cm,width=6cm,angle=-90]{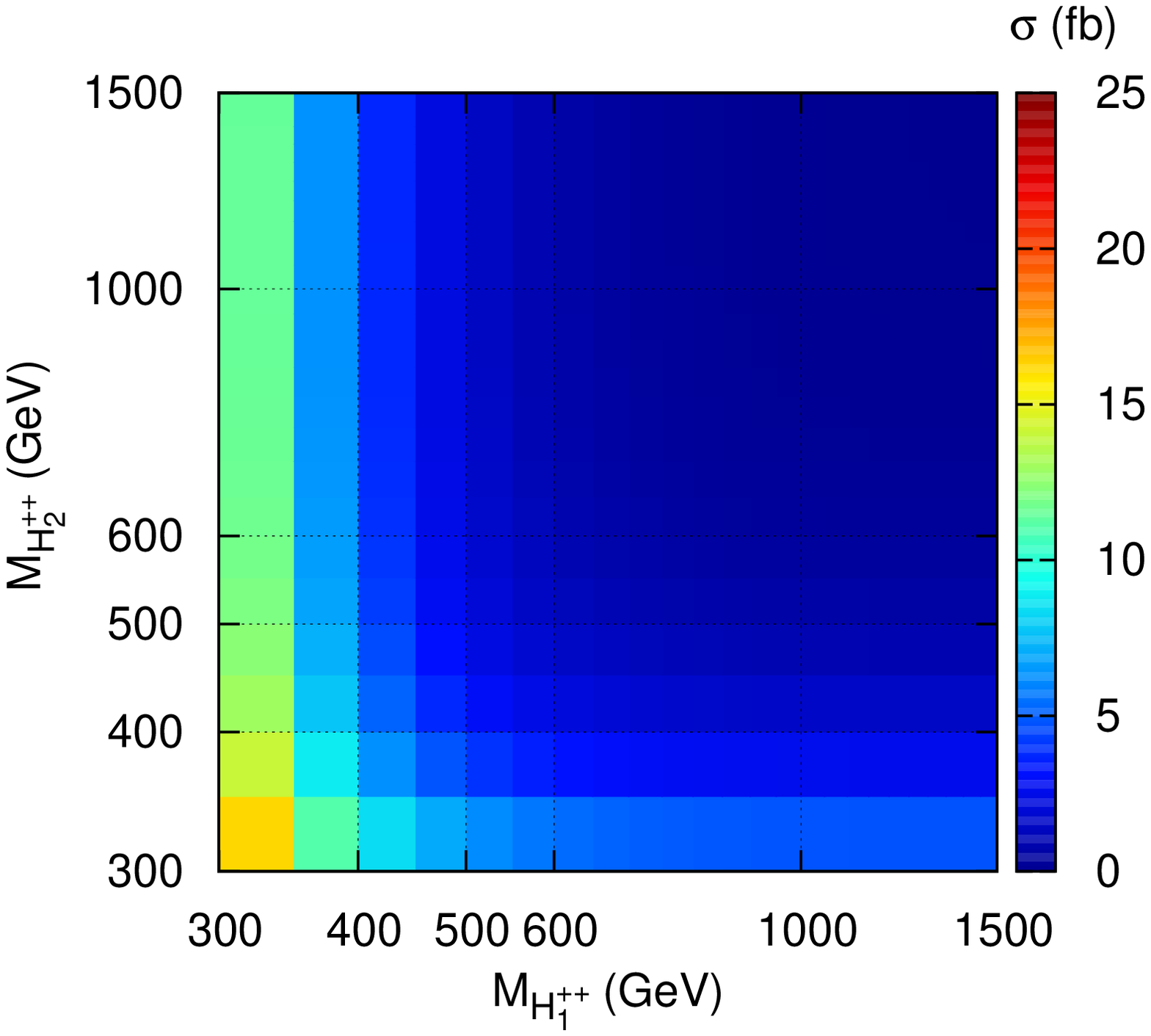}\hspace{-3cm}
\includegraphics[height=10cm,width=6cm,angle=-90]{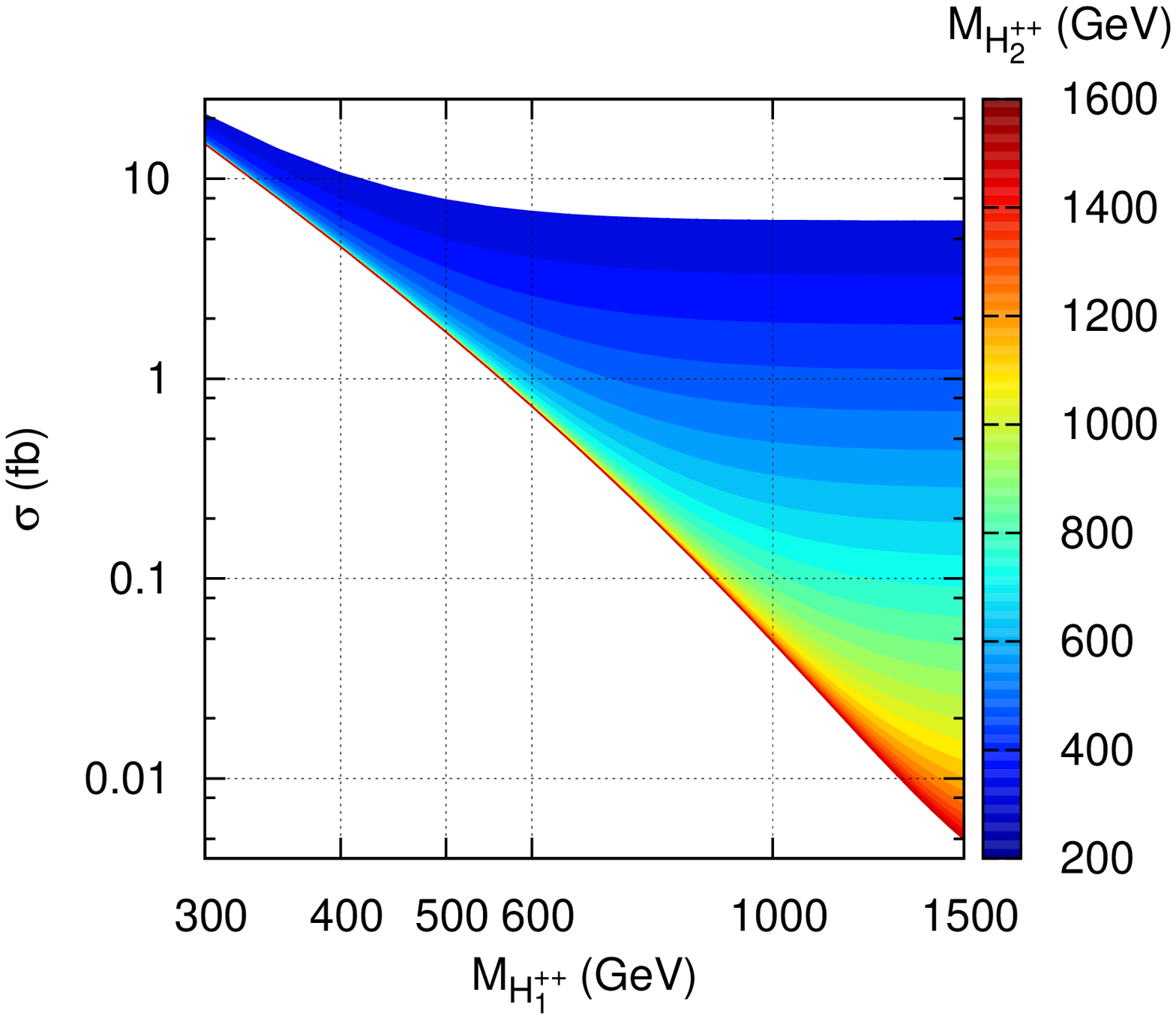}
\caption{Scenario II.  Contour plots for the  $pp \to (H_1^{++} H_1^{--} + H_2^{++} H_2^{--})$ cross section. $\sqrt{s}=14$ TeV, no kinematic cuts imposed.
}
\label{fig:hpphmm_contour}
\end{figure}

\subsubsection*{Scenario II, non-degenerated mass spectrum} 
Here we choose another set of benchmark points where the doubly charged scalars are non-degenerated. 
The cross section for the same process with $M_{H_1^{\pm\pm}} = 400$ GeV and  $M_{H_2^{\pm\pm}} = 500$ GeV at $\sqrt{s}=14$ TeV is 
\begin{equation}
\sigma(pp\rightarrow (H_{1}^{++} H_{1}^{--}+H_{2}^{++} H_{2}^{--})\rightarrow {\ell_i}^+{\ell_i}^+{\ell_j}^-{\ell_j}^-) = 4.95~fb.
\label{eqhpp400500}
\end{equation}
The contributions to the cross sections from individual channels are given as:
\begin{eqnarray}
\sigma(pp\rightarrow H_{1}^{++} H_{1}^{--} ) &=& 1.09 (4.58)~fb, 
\label{eqhpp400500b} \\
\sigma(pp\rightarrow H_{2}^{++} H_{2}^{--} ) &=& 0.13 (0.69)~fb,
\label{eqhpp400500a}
\end{eqnarray}
for $\sqrt{s}=8(14)\;{\rm TeV}$.

Contour plots for the $pp \to (H_1^{++} H_1^{--} + H_2^{++} H_2^{--})$ cross section as a function of doubly charged scalar masses is shown in Fig.~\ref{fig:hpphmm_contour} (left). On the right figure of Fig.~\ref{fig:hpphmm_contour} different projections are used where X and Y axes are for $M_{H_1^{++}}$ and the cross section, respectively, whereas $M_{H_2^{++}}$ is projected as a contour. 
As can be seen from these figures, cross sections at the level of $1~fb$ can be obtained for doubly charged scalar masses  up to
approximately 600 GeV.

\subsubsection{$pp \to H_1^{\pm\pm} H_1^{\mp}$ and $pp \to H_2^{\pm\pm} H_2^{\mp}$} \label{subsubsection:pptoHccHc}

The production of a doubly charged in association with a singly charged
scalar goes through the charged s-channel interaction where  $W_{1,2}^{\pm}$ gauge bosons are exchanged.
Diagrams with s-channel exchanged singly charged scalar $H_2^\pm$ is negligible (its coupling to $W_1$ is proportional to $v_L$ which is zero). As  $W_2^\pm$ is very heavy, the dominant contribution originates from the process via $W_1^\pm$.  
 
To give yet another benchmark, 
we set $v_R=8$ TeV and the following charged scalar masses:
$M_{H_1^{\pm \pm}}$= 483 GeV, $M_{H_2^{\pm \pm}}$ = 527 GeV, $M_{H_1^{\pm}}$ = 355 GeV, $M_{H_2^{\pm}}$ = 15066 GeV. The choice is for the following Higgs potential parameters (for the mass formulas, see the Appendix): $\rho_1=0.2397,\rho_2=0.0005,\rho_3=0.48$,$\lambda_1=0.13,\lambda_2=-0.87,\lambda_3=-5.17$,
$\alpha_3=7.09$.  This example shows that a wide spectrum of charged scalar masses can be easily obtained, still keeping reasonable small potential parameters (important for higher order perturbation analysis).
 To reduce $\tau$ channel decays, the masses for the heavy right handed neutrinos are set at 4 TeV for the first two generations and 800 GeV for the third generation, see Fig.~\ref{bree}.
The cross section for the process before any kinematic cuts
with centre  of mass energy $\sqrt{s} =8(14)$ TeV at the LHC is 
\begin{equation}
\sigma \left(pp\rightarrow (H^{\pm \pm}_{1} H^{\mp}_{1}+H^{\pm \pm}_{2} H^{\mp}_{2})\rightarrow \ell \ell \ell \nu_{\ell}\right) = 1.44(6.05) ~fb.
\label{eqhpphm7423}
\end{equation}
The contributions to the cross sections from individual channels are noted as:
\begin{eqnarray}
\sigma (pp\rightarrow H^{\pm \pm}_{1} H^{\mp}_{1} ) &=& 1.48 (6.24)~fb, 
\label{eqhpphm7423a} \\
\sigma (pp\rightarrow H^{\pm \pm}_{2} H^{\mp}_{2} ) &\sim& 0 (0)~fb,
\label{eqhpphm7423b} 
\end{eqnarray}
with $\sqrt{s}=8(14)\;{\rm TeV}$.

For the model consistency (i.e. chosen potential parameters), the second singly charged scalar has been chosen with very high mass 
$M_{H_2^{\pm}}=15066$ GeV.
Even if it has low mass ($\sim 400$ GeV) then also the cross section for the processes 
$pp\xrightarrow{} H^{\pm \pm}_{2} H^{\mp}_{2}$  is very low compared 
to $pp\xrightarrow{} H^{\pm \pm}_{1} H^{\mp}_{1}$ as $H^{\pm \pm}_{2} H^{\mp}_{2} W_{1}^{\mp}$  
coupling is proportional to $\sin{\xi}$ and $H^{\pm \pm}_{2} H^{\mp}_{2} W_{2}^{\mp}$  coupling is proportional to $\cos{\xi}$. 
On the other hand, $H^{\pm \pm}_{1} H^{\mp}_{1} W_{1}^{\mp}$  coupling is proportional to $\cos{\xi}$ and $H^{\pm \pm}_{1} H^{\mp}_{1} W_{2}^{\mp}$ 
coupling is proportional to  $\sin{\xi}$. In both cases $W_2^{\pm}$ mediated processes are much less dominant than the $W_1^{\pm}$ mediated processes. 
But as the charged gauge boson mixing angle $\xi$ is neglected, the $H^{\pm \pm}_{2} H^{\mp}_{2} W_{1}^{\mp}$ vertex is much more suppressed
compare to $H^{\pm \pm}_{1} H^{\mp}_{1} W_{1}^{\mp}$. 

\begin{figure}[h]
\begin{center}
\includegraphics[width=7cm,angle=-90]{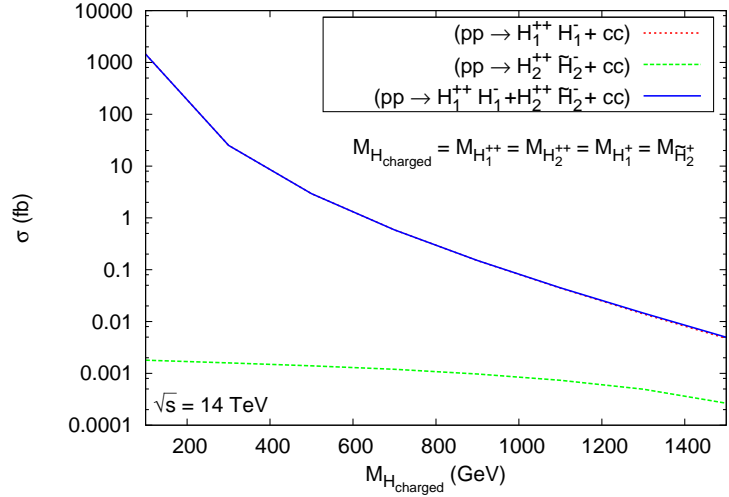}
\end{center}
\caption{Production cross sections for $pp \to H_1^{++} H_1^{-}$ and $pp \to H_2^{++} \tilde{H}_2^{-}$ processes at $\sqrt{s}=14$ TeV and no kinematic cuts are imposed. Mass of $H_2^\pm$ is allowed to be small and denoted with a tilde.}
\label{fig:crosssection_pptoHcchc}
\end{figure}

\begin{figure}[h]
\hspace{-0.4cm}
\begin{center}
\includegraphics[width=7cm,angle=-90]{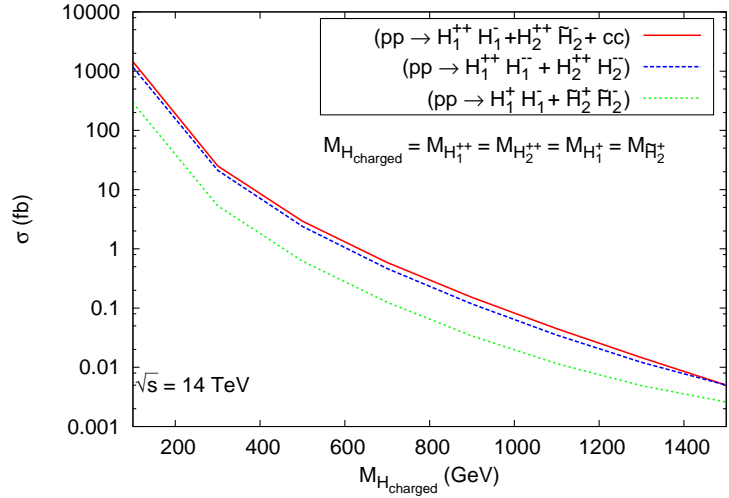}
\end{center}
\caption{Summary of various MLRSM LHC production cross sections considered in the
paper is shown with charged scalars at $\sqrt{s}=14$ TeV and without kinematic cuts. We have taken degenerate mass $M_{H_{charged}}$ for  $M_{H_1^{++}}$, $M_{H_2^{++}}$, $M_{H_1^{+}}$ and $M_{\tilde{H}_2^{+}}$.}
\label{fig:crosssection_pptochargedscalars}
\end{figure}  
It appears that in MLRSM mixed processes, $pp \to H_1^{++} H_2^{-}$ and $pp \to H_2^{++} H_1^{-}$, vanishes as $v_L=0$.
In Fig.~\ref{fig:crosssection_pptoHcchc} the total cross section for two considered processes are given.
The mass of  $H_2^\pm$ is allowed to be small and because, as discussed before, this is not natural in the MLRSM, its contribution is denoted with a tilde.
Anyway, its contribution (keeping a form of its couplings as dictated by MLRSM) is negligible. 
Final comparison of cross sections of different processes discussed in sections \ref{subsubsection:pptoHmHp}, \ref{subsubsection:pptoHmmHpp} and \ref{subsubsection:pptoHccHc}
is given in Fig.~\ref{fig:crosssection_pptochargedscalars}. We can see that the largest cross sections are for a pair production of singly with doubly charged scalars, and 
the cross sections for production of doubly charged scalar pair is slightly lower, while the smallest cross section is for pair production of singly charged scalars. Contributions from processes where $H_2^\pm$ is involved are negligible or at most much smaller than corresponding results where $H_1^\pm$ is involved.  
Keeping in mind the status of the SM background (analysed for our purposes in section~\ref{sec:background_signal_significance}) we look for multi lepton signals for three or more leptons. 
Thus we focus in the following sections on the processes which involve primary production of at least one doubly charged scalar.  

\subsection{Primary production of a heavy Higgs and gauge bosons}

\subsubsection{$pp \to  W_{1/2}^{\mp} H_{1/2}^\pm $, $pp \to Z_{1/2} H_{1/2}^{\pm}$ and $pp \to \gamma H_{1/2}^{\pm}$}

In our scenarios the production cross sections for these processes are very small and can be ignored. 
This is because the $W_2/Z_2$ propagator diagrams are suppressed as they are as heavy as few TeV. For the other light propagators 
the scalar-gauge boson-gauge boson vertices are proportional to $\sin \xi$ and/or $v_L$, which are zero here.

\section{Simulations and results for final lepton signals \label{sim} } 

In this paper we are interested in tri- and four-lepton signal events. 
To enhance such signals, suitable kinematic cuts are applied in order to decrease the SM backgrounds. 

\subsection{Events selection criteria} \label{sec:event_selection}

The detailed simulation criteria  used in our study are following:
\begin{itemize}
 \item The Parton Distribution Function (PDF): CTEQ6L1 \cite{Pumplin:2002vw}.
\item Initial selection (identification) criteria of a lepton: pseudorapidity $|\eta| < 2.5$ and $p_T$ (transverse momentum $p_T=\sqrt{{p_x}^2 + {p_y}^2}$) of that lepton should be $>$ 10 GeV.
\item Detector efficiency for leptons: 
\begin{itemize}
\item[$\diamondsuit$] For electron (either $e^-$ or $e^+$) detector efficiency is 0.7 ($70\%$);
\item[$\diamondsuit$] For muon (either $\mu^-$ or $\mu^+$) detector efficiency is 0.9 ($90\%$).
\end{itemize}
\item Smearing of electron energy and muon $p_T$ are considered. All these criteria are implemented in PYTHIA  and for details see \cite{gb-jd-sg-pk}.
\item Lepton-lepton separation: The separation between any two leptons should be $\Delta R_{ll} \ge 0.2$. 
\item Lepton-photon separation: $\Delta R_{l\gamma} \ge 0.2$ with all the photons having ${p_T}_\gamma > 10$ GeV.
\item Lepton-jet separation: The separation of a lepton with all the jets should be $\Delta R_{lj} \ge 0.4$, otherwise that lepton is not counted as lepton. 
Jets are constructed from hadrons using PYCELL within the PYTHIA.
\item Hadronic activity cut: This cut is applied to take only pure kind of leptons that have very less hadronic activity around them.  
Each lepton should have hadronic activity, $\frac{\sum p_{T_{hadron}}}{p_{T_l}} \le 0.2$ within the cone of radius 0.2 around the lepton. 
\item Hard $p_T$ cuts: ${p_T}_{l_1}>30$ GeV, ${p_T}_{l_2}>30$ GeV, ${p_T}_{l_3}>20$ GeV, ${p_T}_{l_4}>20$ GeV. 
\item Missing $p_T$ cut: This cut is not applied for four-lepton final states while for three-lepton case due to the presence of 
neutrino, a missing  $p_T$ cut ($ > 30$ GeV) is applied.
\item Z-veto\footnote{Same flavoured but opposite sign lepton pair invariant mass $m_{\ell_1
\ell_2}$ must be sufficiently away from $Z_1$ mass, such that, typically,  $|m_{\ell_1\ell_2} - M_{Z_1}| \geq 6 \Gamma_{Z_1} \sim 15$
GeV  \cite{gb-jd-sg-pk}.} is also applied to suppress the SM background. This has larger impact while reducing the background for four-lepton without missing energy.

\end{itemize}

\subsection{Signal events for doubly charged Higgs particles in MLRSM}

Doubly charged scalars decay mainly to either a pair of same sign charged leptons or charged gauge bosons depending on the choice of parameters. 
As already discussed, we have chosen the parameter space in such a way that the doubly charged scalars decay to charged leptons with almost 100\% branching ratio. 
 
This decay is lepton number violating and can also be possibly lepton flavour violating. In our scenarios we assume no lepton flavour violation as the Yukawa
couplings are considered to be flavour diagonal. Thus, the four lepton final state contains two pairs of same sign and same flavoured charged leptons where each pair has opposite charges to each other. As there is no neutrino (missing energy) or jet involved it is easy to reconstruct the momentum of the final state particles. We have reconstructed invariant masses\footnote{The invariant mass for a lepton pair is defined as $m_{\ell_1\ell_2} = \sqrt{(E_1 + E_2)^2 - (\vec{P_1} + \vec{P_2})^2}
$, where $E_i$ and $\vec{P_i}$ are the energy and three
momentum of $\ell_i$, respectively.} for same sign di-leptons (SSDL) and opposite sign di-leptons (OSDL). As the doubly charged scalars are the parents of the di-lepton pairs,
invariant mass of the SSDL is expected to give a clean peak around the mass of the doubly charged scalar, which is not necessarily a case for OSDL.

\subsubsection{$pp \to H_1^{++} H_1^{--}$ and $pp \to H_2^{++} H_2^{--}$}

\subsubsection*{Scenario I, degenerated doubly charged mass spectrum}
\begin{figure}[h!]
\hspace{-0.4cm}
\begin{center}
\includegraphics[width=7cm,angle=-90]{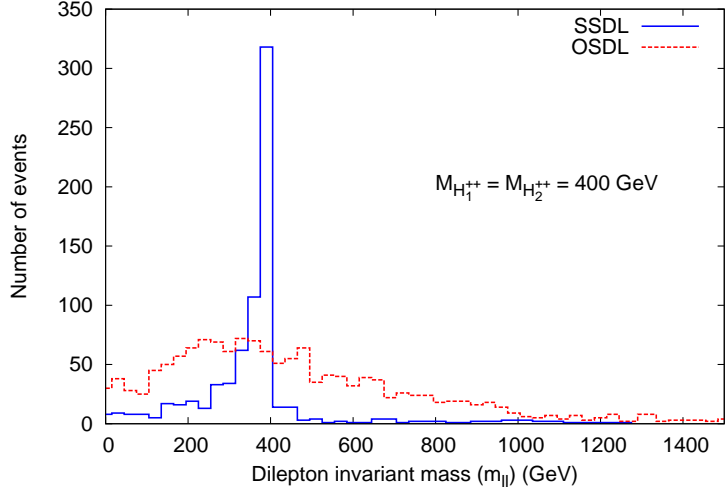}
\end{center}
\caption{Invariant mass for SSDL and OSDL for $(pp\rightarrow H_{1,2}^{++} H_{1,2}^{--} \to 4 l)$ with $M_{H_1^{++}} = M_{H_2^{++}} = 400$ GeV for  $\sqrt{s}=14$ TeV and $L=300 ~fb^{-1}$. As the doubly charged scalars are degenerate in mass both the invariant mass peaks occur at the same place and thus cannot be distinguished.}
\label{fig:pptoHppHmmtollll}
\end{figure}

As calculated in Section II, Eq.~(\ref{eqhpp814}), if $M_{H_1^{++}} = M_{H_2^{++}} = 400$ GeV, the cross section at the LHC with centre of mass energy $\sqrt{s}=14$ TeV  is 
$\sigma(pp\rightarrow (H_{1}^{++} H_{1}^{--}+H_{2}^{++} H_{2}^{--})\rightarrow {\ell_i}^+{\ell_i}^+{\ell_j}^-{\ell_j}^-) = 6.06~fb$, where $\ell_{i,j} = e,\mu$.  
After implementing all the cuts, as described in section~\ref{sec:event_selection}, the four lepton events with no missing energy can be estimated. 
 Each pair of SSDL originates from different doubly charged 
scalars. We have plotted the reconstructed invariant mass distributions for 
both SSDL and OSDL in Fig.~\ref{fig:pptoHppHmmtollll} with anticipated integrated luminosity $L=300 ~fb^{-1}$. As both the doubly charged scalars are degenerate the invariant mass peaks occur 
at around 400 GeV. \emph{This clean reconstruction of the invariant mass is indeed possible even in the hadronic environment and can be a smoking gun feature indicating the presence of doubly charged scalars.}

We have computed this process also with centre of mass energy 8 TeV.
In this case we find that the cross section, with $\sqrt{s} = 8$ TeV at the LHC, is 1.06 $fb$, about 6 times smaller than for $\sqrt{s}=14$ TeV.
 If we take 
present integrated luminosity to be $25~fb^{-1}$ then total number of the events even before all the cuts, is statistically 
insignificant to analyse this particular process at the LHC after implementing all the selection criteria. 
Thus to justify this four lepton signal for this scenario needs more data in future.

\begin{figure}[htb]
\hspace{-0.4cm}
\begin{center}
\includegraphics[width=7cm,angle=-90]{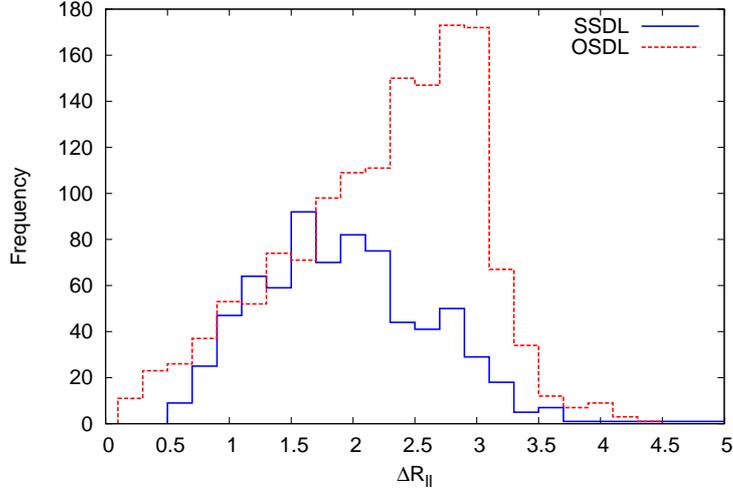}
\end{center}
\caption{Lepton - lepton separations for the same sign lepton pairs (\textcolor{blue}{$\Delta R_{\ell^{\pm} \ell^{\pm}}$}) and opposite sign lepton pairs 
(\textcolor{red}{$\Delta R_{\ell^{\pm} \ell^{\mp}}$}) for $(pp\rightarrow H_{1,2}^{++} H_{1,2}^{--} \to 4 l)$ within the degenerate scenario
 with $M_{H_1^{++}} = M_{H_2^{++}} = 400$ GeV for  $\sqrt{s}=14$ TeV and $L=300 ~fb^{-1}$. }
\label{fig:deltaRll_setI}
\end{figure}

To select the doubly charged scalar signal properly and in an independent way, there is another interesting variable which can be used for determination of 
signals as suggested in \cite{Babu-Ayon} 
\begin{equation}
\Delta R_{\ell_1 \ell_2} = \sqrt{(\eta_1 - \eta_2)^2 + (\phi_1 - \phi_2)^2},
\label{rll}
\end{equation}
 where $\eta_i$ and $\phi_i$ denote pseudorapidity and azimuth of $\ell_i$, respectively.
$\Delta R_{\ell \ell}$ amounts the separation between two light charged leptons ($\ell$) in azimuth-pseudorapidity plane. Its physical importance is that in the detector
if $\Delta R_{\ell \ell}$ is smaller than the specified value then one can not distinguish whether the  deposited energy is really by one or two leptons.
So, one chooses only events for which leptons are well separated.
\emph{We expect that the leptons originated from a single doubly charged scalar will be less separated than the leptons coming from different charged scalars.} In our considered 
processes and decays the doubly charged scalars decay mainly into pair of same flavoured same sign leptons. Thus  in a case of opposite sign di-lepton pair 
each of them are coming from different doubly charged scalars must be well separated. We have plotted the $\Delta R_{\ell \ell}$ distribution to 
address this feature. It is pretty clear from Fig.~\ref{fig:deltaRll_setI} that the distribution peaks at smaller $\Delta R_{\ell \ell}$ 
for same sign lepton pair while that for the oppositely charged lepton pair peaks at larger value of $\Delta R_{\ell \ell}$, as expected. This implies that most of the leptons in the SSDL pairs are less separated than the leptons which belong to the OSDL pair.

\subsubsection*{Scenario II, non degenerated doubly charged mass spectrum}
\begin{figure}[htb]
\hspace{-0.4cm}
\begin{center}
\includegraphics[width=7cm,angle=-90]{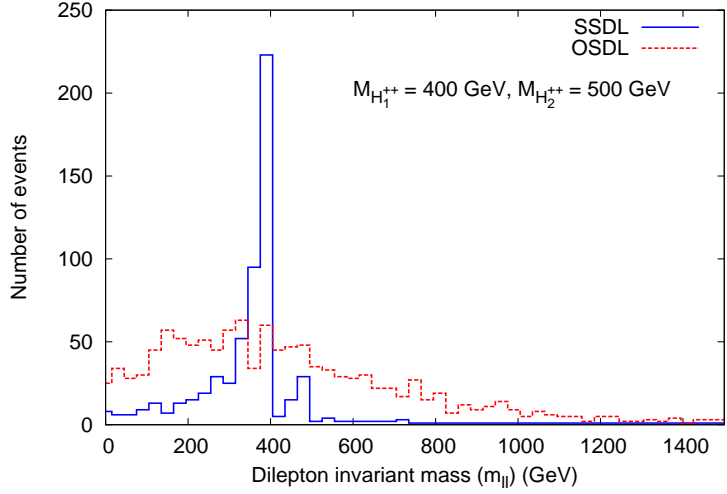}
\end{center}
\caption{Invariant mass for SSDL and OSDL signals in the $(pp\rightarrow H_{1,2}^{++} H_{1,2}^{--} \to 4 l)$ process in the non-degenerate mass scenario with $M_{H_1^{\pm\pm}} = 400$ GeV and  $M_{H_2^{\pm\pm}} = 500$ GeV for $\sqrt{s}=14$ TeV and $L=300 ~fb^{-1}$.}
\label{fig:pptoHppHmmtollll_500}
\end{figure}

Here we choose another set of benchmark points where the doubly charged scalars are non-degenerate. 
In Section II, Eq.~(\ref{eqhpp400500}), the cross section at $\sqrt{s}=14$ TeV has been calculated for the same process with $M_{H_1^{\pm\pm}} = 400$ GeV and  $M_{H_2^{\pm\pm}} = 500$ GeV, $\sigma=4.95~fb$.
As $M_{H_2^{\pm \pm}}>M_{H_1^{\pm \pm}}$, the production cross section for $H_1^{\pm \pm}$ is much larger than that for $H_2^{\pm \pm}$.
\emph{Thus the four lepton events will be generated mostly from the leptonic decays of the $H_1^{\pm \pm}$ pair than $H_2^{\pm \pm}$ decays.}
This statement is very 
distinctively clear from the invariant mass distributions of the same sign di-leptons, as shown in the Fig.~\ref{fig:pptoHppHmmtollll_500}.
Maximum number of same di-lepton events are with an invariant mass peak around $M_{H_1^{\pm\pm}}=400$ GeV and that around 
$M_{H_2^{\pm\pm}}=500$ GeV is much smaller, as expected.

\begin{figure}[hb]
\hspace{-0.4cm}
\begin{center}
\includegraphics[width=7cm,angle=-90]{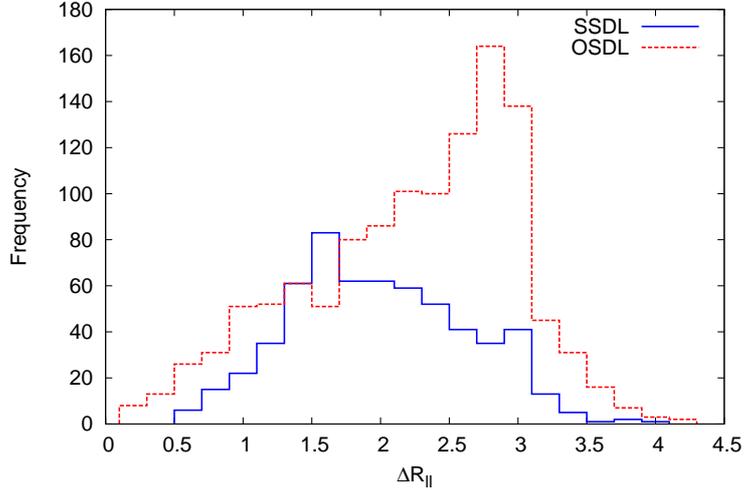}
\end{center}
\caption{Lepton - lepton separations for same sign lepton pairs (\textcolor{blue}{$\Delta R_{\ell^{\pm} \ell^{\pm}}$}) and opposite sign lepton pairs (\textcolor{red}{$\Delta R_{\ell^{\pm} \ell^{\mp}}$}) in the $(pp\rightarrow H_{1,2}^{++} H_{1,2}^{--} \to 4 l)$ process for non-degenerate mass scenario having $M_{H_1^{\pm\pm}} = 400$ GeV and  $M_{H_2^{\pm\pm}} = 500$ GeV with $\sqrt{s}=14$ TeV and $L=300 ~fb^{-1}$. }
\label{fig:deltaRll_setII}
\end{figure}

We also performed  the $\Delta R_{\ell \ell}$ distribution for the same benchmark point. For the same reason as explained before 
our expectation is reflected in Fig.~\ref{fig:deltaRll_setII}.

\subsubsection{$pp \to H_1^{\pm \pm} H_1^{\mp}$ and $pp \to H_2^{\pm \pm} H_2^{\mp}$}
 These processes lead to the tri-lepton events with missing $p_T$, see Table~\ref{tabrange1}. 
For chosen MLRSM parameters, Eq.~(\ref{eqhpphm7423}), 
the cross section for the process $pp\rightarrow (H^{\pm \pm}_{1} H^{\mp}_{1} + H^{\pm \pm}_{2} H^{\mp}_{2}) \rightarrow \ell \ell \ell \nu_{\ell}$ before cuts
with centre of mass energy $\sqrt{s} = 14$ TeV   is $\sigma = 6.05 ~fb$. The tri-lepton events can be classified into two categories: either $\ell^+ \ell^+ \ell^-$ or $\ell^- \ell^- \ell^+$.
The first and second types of signals are originated from $W_1^+$ and $W_1^-$ mediated processes, respectively. Thus, it is indeed possible to estimate the charge asymmetry, 
define as the ratio of the number of events of $\ell^+ \ell^+ \ell^-$ type to the number of events of $\ell^- \ell^- \ell^+$ type at the LHC. This is very similar 
to the forward-backward asymmetry at Tevatron. 
This charge asymmetry depends on Parton Distribution Functions (PDF) and thus is a special feature of LHC. We have estimated this ratio ($R^{+}_{-}$) 
with the above choices of charged scalar masses with $\sqrt{s} = 14$ TeV and integrated luminosity 300 $fb^{-1}$. 
We find 554 tri-lepton signal events after all the cuts and that leads to
\begin{equation} 
R^{+}_{-}=\frac{\#~ \mbox{of events for} ~\ell^+ \ell^+ \ell^-}{\# ~\mbox{of events for} ~\ell^- \ell^- \ell^+} = \frac{396}{158} \simeq 2.51.
\label{rpm}
\end{equation} 

In SM the corresponding value calculated for the main processes given in the next section in Table~\ref{table:background} is
 $(R^{+}_{-})_{SM} = \frac{17.751}{14.962} = 1.186$. This value is slightly different from the calculated values in \cite{Kom:2010mv} where higher order corrections are taken into account and the specific kinematic cuts are different. Nevertheless, MLRSM value given in Eq.~(\ref{rpm})
 differs substantially from its SM counterpart to signify its presence.

\begin{figure}[t]
\hspace{-0.4cm}
\begin{center}
\includegraphics[width=7cm,angle=-90]{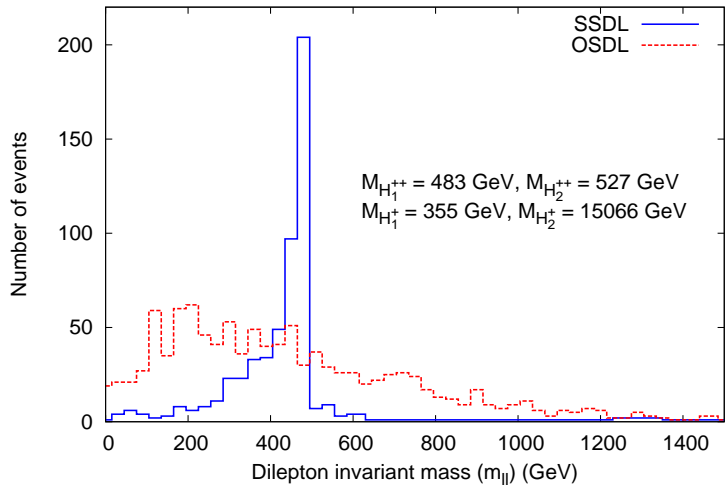}
\end{center}
\caption{Invariant mass plots for SSDL and OSDL for the signals $\ell^{\pm} \ell^{\pm} \ell^{\mp}$ + missing $p_T$, 
at the LHC with $\sqrt{s} = 14$ TeV and integrated luminosity 300 $fb^{-1}$.}
\label{fig:Inv_mass_HdcHsc_cteq6l1}
\end{figure}

As discussed in Section II, the $H^{\pm \pm}_{2} H^{\mp}_{2} W_{1}^{\mp}$ vertex is much more suppressed
compare to $H^{\pm \pm}_{1} H^{\mp}_{1} W_{1}^{\mp}$. Thus, in this case most of the tri-lepton events are originated from $pp\xrightarrow{} H^{\pm \pm}_{1} H^{\mp}_{1}$ process.  This is clearly visible from the invariant mass distributions. Here we have plotted the same and opposite sign di-lepton invariant mass distributions, 
see Fig.~\ref{fig:Inv_mass_HdcHsc_cteq6l1}. As similar to the earlier discussions \emph{in the opposite sign lepton pairs two leptons have different origin 
thus their invariant mass distribution is continuous while the same sign di-lepton invariant mass distributions always peak around the mass of the doubly charged scalars}.

Here, from Fig.~\ref{fig:Inv_mass_HdcHsc_cteq6l1}, it is distinctly seen that the significant amount of same sign di-lepton pair peaks at 
$M_{H_1^{\pm \pm}}=483$ GeV rather than $M_{H_2^{\pm \pm}}=527$ GeV. This implies that the dominant contribution to this tri-lepton events are generated through
$pp \to H_1^{\pm \pm} H_1^{\mp}$ process (cf. Fig.~\ref{fig:crosssection_pptoHcchc}) and the further leptonic decays of the charged scalars.

\begin{figure}[h]
\hspace{-0.4cm}
\begin{center}
\includegraphics[width=7cm,angle=-90]{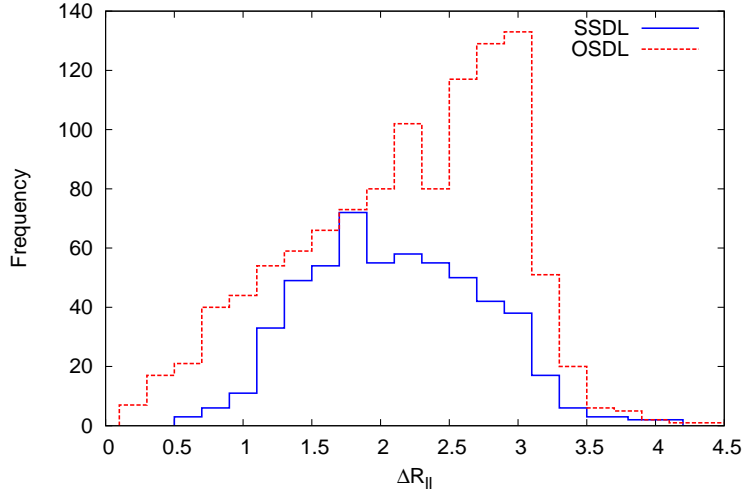}
\end{center}
\caption{Lepton-lepton separation plot for same sign leptons (\textcolor{blue}{$\Delta R_{\ell^{\pm} \ell^{\pm}}$}) 
and opposite sign leptons (\textcolor{red}{$\Delta R_{\ell^{\pm} \ell^{\mp}}$}) in the process $(pp\rightarrow (H^{\pm \pm}_{1} H^{\mp}_{1} + H^{\pm \pm}_{2} H^{\mp}_{2}) \rightarrow 3 \ell )$. Here $\sqrt{s} = 14$ TeV and 
integrated luminosity 300 $fb^{-1}$ at the LHC. }
\label{fig:Deltarll_setIII}
\end{figure}

In the Fig.~\ref{fig:Deltarll_setIII}, separations between leptons are plotted. As can be seen from this figure
the SSDL separations peak at lower value of $\Delta R_{\ell \ell}$, while OSDL separations peak at larger value of
$\Delta R_{\ell \ell}$. This is because same-sign leptons pair has the origin from the same mother, while opposite sign leptons pair
has both the leptons from different mothers.

For $\sqrt{s} = 8$ TeV and the same benchmark point  the production cross section
$\sigma(pp\rightarrow (H^{\pm \pm}_{1} H^{\mp}_{1} + H^{\pm \pm}_{2} H^{\mp}_{2}) \rightarrow \ell \ell \ell \nu_{\ell}) = 1.44~ fb$ is about four times smaller than for $\sqrt{s} = 14$ TeV,
Eq.~(\ref{eqhpphm7423}).
With an integrated luminosity 25 $fb^{-1}$ at $\sqrt{s} = 8$ TeV and 300 $fb^{-1}$ at $\sqrt{s} = 14$ TeV, total number of events is about 50 times smaller in the former case, so the difference is substantial. 

Distributions presented so far show that it is possible to extract clear signals for doubly charged scalars at the LHC. However, for signal identification crucial is 
how large the SM background effects are and the significance too.

\subsection{Background estimation and Significance of signals} \label{sec:background_signal_significance}

Kinematic cuts are used which have been investigated and established for the first time in  \cite{gb-jd-sg-pk}. The cuts are optimised 
in a way such that we can reduce the SM background and enhance the signal events\footnote{In our analysis while computing the tri-lepton events (signal and background), the $p_T$ of the third hardest lepton needs to be greater than 20 GeV, and also a missing $p_T$ cut ($>$ 30 GeV) must be satisfied, see section 
\ref{sec:event_selection}. Thus the tri-lepton background for process like $t\bar{t}$ where one of the lepton is coming from semi-leptonic decays of $B$'s is reduced. Here the hadronic activity cut also reduces the hadronic activity around the selected leptons and plays a crucial role in this case. All these cuts reduces the efficiency of misidentification of b-jets as leptons. In our case this is less than 0.05\%.}.
Standard Model background cross sections for tri- and four-lepton signals are given in Table~\ref{table:background}. In this table we have also separately computed the backgrounds for 
$\ell^+\ell^+\ell^-$ and $\ell^-\ell^-\ell^+$.
\begin{table}[h!]
\begin{center}
\begin{tabular}{|l|c|c|c|c|} \hline
\bf{processes}   & $\bf{3 \ell}$ ($fb$) & $\ell^+\ell^+\ell^-$ ($fb$) & $\ell^-\ell^-\ell^+$ ($fb$) & $\bf{4\ell}$ ($fb$)\\ \hline
$t \bar{t}$                                & $18.973$& $9.522$&$9.451$  &   --                        \\    \hline
$t \bar{t} (Z / \gamma^{\star})$         & $1.103$   &$0.549$&$0.552$ & $0.0816$  \\    \hline
$t \bar{t} W^\pm$                          & $0.639$ &$0.422$ &$0.214$& --      \\    \hline
$W^\pm (Z / \gamma^{\star})$             & $10.832$ & $6.664$&$4.164$&   --                       \\    \hline
$(Z/\gamma^{\star})(Z/\gamma^{\star})$ &    $1.175$     &$0.594$&$0.581$& $0.0362$             \\    \hline \hline
    \bf{TOTAL}                             & $\bf{32.722}$ &$\bf{17.751}$& $\bf{14.962}$& $\bf{0.1178}$     \\    \hline
\end{tabular}
\end{center}
\caption{Dominant Standard Model background contributions (in $fb$) for tri- and four-lepton signals at the LHC with $\sqrt{s} = 14$ TeV after obeying suitable
selection criteria defined in the text.  The $t \bar{t}$ cross section is presented here after the inclusion of k-factor. While computing the SM contributions to $4\ell$ final state, no missing $p_T$ cut has been applied.}
\label{table:background}
\end{table}

In principal the tri-lepton contributions can come also from $H_1^{++}H_1^{--}$  and $H_2^{++}H_2^{--}$ involved processes if during simulations one of the four-leptons does not satisfy the cuts. But in our case this contribution is negligible due to the extra missing energy cut applied as one of  the gate pass for the tri-lepton events. Thus all the productions together are considered and all the intermediate particles are allowed to decay. After passing through the cuts, tri-lepton and four-lepton events are counted.

\begin{table}[htb]\label{table:signals}
\begin{center}
\begin{tabular}{|c|c|c|c|c|c|}
\hline
\multirow{2}{*}{\textbf{Luminosity}} & Background & Signal & Background & \multicolumn{2}{|c|}{Signal {$4 \ell$} events  } \\ 
                            &        $3 \ell$  events   &      $3 \ell$   events &      {$4 \ell$} events        & scenario I         & scenario II        \\ \hline
25   $fb^{-1}$                          & 797.5      & 46.2   & 2.9        & (i) 30             & 24.8               \\
                            &            &        &            & (ii) 4.4           &                    \\ \hline
300  $fb^{-1}$                          & 9569.7     & 554    & 34.8       & (i) 360            & 298                \\
                            &            &        &            & (ii) 53            &                    \\ \hline
\end{tabular}

\end{center}
\caption{\textcolor{black}{Number of background and signal events at 25 $fb^{-1}$ and 300 $fb^{-1}$ as an anticipated integrated luminosity at next 14 TeV run of LHC. The tri-lepton signal is computed for following charged scalar masses: $M_{H_1^{\pm \pm}}$= 483 GeV, $M_{H_2^{\pm \pm}}$ = 527 GeV, $M_{H_1^{\pm}}$ = 355 GeV, $M_{H_2^{\pm}}$ = 15066 GeV.
Scenario I reflects degeneracy of doubly charged scalar masses with (i) $M_{H_1^{\pm \pm}}=M_{H_2^{\pm \pm}}=400$  GeV and (ii) $M_{H_1^{\pm \pm}}=M_{H_2^{\pm \pm}}=600$ GeV, while Scenario II realises their non-degenerate spectrum, namely 
$M_{H_1^{\pm \pm}}=400$    and $M_{H_2^{\pm \pm}}=500$ GeV. Here we have used the same kinematical cuts as applied while estimating the SM background events. We have not implemented other extra cuts, like invariant mass ($m_{\ell \ell}$) and lepton separation ($\Delta R_{\ell \ell}$) to estimate the signal and background events in this Table.}}
\label{table:background_signal_events}
\end{table}

\begin{table}[htb]\label{table:significance}
\begin{center}
\begin{tabular}{|c|c|c|c|c|}
\hline
Significance &  $3 \ell$   events  & \multicolumn{2}{|c|}{{$4 \ell$} events  } \\ 
        &                               & scenario I         & scenario II        \\ \hline
S/$\sqrt{B}$                   &  5.66          & (i) NA          & NA              \\
              &                                          & (ii) NA           &                    \\ \hline
S/$\sqrt{(S+B)}$                    & 5.51         & (i) 18.11         & 16.34               \\
              &                                   & (ii) 5.65           &                    \\ \hline
\end{tabular}

\end{center}
\caption{The significance of the signals given in Table~\ref{table:background_signal_events} is given using two
definitions of significance: (i) $S/\sqrt{B}$, and (ii) $S/\sqrt{(S+B)}$, where $S$ and $B$ are the total number of signal and background events for 300$~fb^{-1}$ integrated luminosity, respectively.
The parameters are the same as given in Table~\ref{table:background_signal_events}. Here `NA' implies that S/$\sqrt{B}$ can not be used as the definition of significance in these cases as S$<<B$ is not justified.}
\label{table:significance}
\end{table}

In Table~\ref{table:background_signal_events} we present the total background and signal events for 25 and 300$~fb^{-1}$ integrated luminosities. It is clear that four-lepton signals are well beyond the SM background. The tri-lepton signal is also very prominent over the background (what
matters is the signal excess over the background fluctuations).
To see it properly, in Table~\ref{table:significance}  the significance of different signals
is shown.

Assuming the significance at the level of 5 as a comfortable discovery limit, we can see that LHC will be in the next run sensitive to masses of MLRSM doubly charged Higgs bosons up to approximately 600 GeV.

\section{MLRSM charged Higgs bosons contribution to $H_0^0 \to \gamma \gamma$}   
\begin{figure}[h!]
\hspace{-0.4cm}
\begin{center}
\includegraphics[width=9cm,angle=0]{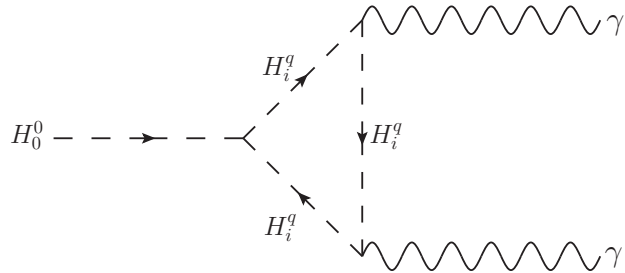}
\end{center}
\caption{Charged scalar contribution to the $H_0^0 \to \gamma \gamma$ channel at the LHC. 
In the loop there are three contributions coming from the charged scalars $H_i^q\equiv H_{1}^{\pm \pm},H_{2}^{\pm \pm}, H_{1}^{\pm}$.  
In MLRSM $H_{2}^{\pm}$ is very heavy and its contribution is negligible.}
\label{fig:Htogammagamma_diagram}
\end{figure}

In LR symmetric models there are (singly-, doubly-) charged scalars and charged gauge boson ($W_2^{\pm}$) 
which couple to photons and hence they can contribute to $H_0^0 \to \gamma \gamma$ channel where $H_0^0$ is the SM-like neutral Higgs taken to be 125 GeV. Since $W_2^{\pm}$ are heavy, their contributions are 
suppressed compared to charged scalars, so we look for charged scalar contributions. They contribute to the channel via a loop shown in the 
Fig.~ \ref{fig:Htogammagamma_diagram}. 

Following  \cite{PhysRevD.40.1722,Picek:2012ei,Carena:2012xa}  we can write the enhancement factor for this channel, 
which is nothing but a ratio of partial decay width in the new model to that 
in the SM

\begin{equation}
 R_{\gamma\gamma} =  \left|1 + \sum_{S=H_{1,2}^{\pm \pm},H_{1}^{\pm}} Q_S^2 \frac{c_S}{2} \frac{k_+^2}{M_S^2} 
 \frac{A_0(\tau_S)}{A_1(\tau_{W_1}) + N_c Q_t^2 A_{1/2}(\tau_t)} \right|^2.
\end{equation}
In the above equation $Q_S$ is electric charge of charged scalars in unit of $e$, $M_S$ is a mass of scalars. $N_c$ is colour factor which is 1 for colour 
singlet scalars and $\tau_{i} = 4 m_i^2/m_{H_0^0}^2 (i = W_{1},t,S)$.
$c_S$ are the coupling of the Higgs boson with the charged scalars and $k_+ = \sqrt{k_1^2 + k_2^2}$ where $k_1,k_2$ are the vacuum expectation values of the bi-doublet.
The expressions for $c_S$ are as follows
\begin{eqnarray}
c_{H_0^0H_1^+H_1^-} =-\Biggl[\frac{2 {\alpha_1} k_+^2+8 {\alpha_2} {k_1} {k_2}+{\alpha_3} (k_+^2)}{2 {k_+^2}}\Biggr] ,\label{cs1}\\
c_{H_0^0H_1^{++}H_1^{--}} =-\Biggl[\frac{{\alpha_1} k_+^2+{k_1} (4 {\alpha_2} {k_2}+{\alpha_3} {k_1})}{{k_+^2} }\Biggr], \label{cs2}\\
c_{H_0^0H_2^{++}H_2^{--}} =-\Biggl[\frac{{\alpha_1} k_+^2+{k_1} (4 {\alpha_2} {k_2}+{\alpha_3} {k_1})}{{k_+^2} }\Biggr] \label{cs3} .
\end{eqnarray}
Here the parameters that are involved in the above Eqs.~(\ref{cs1}-\ref{cs3}), are contained in the scalar potential and following the convention as suggested in \cite{Duka:1999uc}.

\begin{figure}[htb]
\hspace{-0.4cm}
\begin{center}
\includegraphics[height=8cm,angle=0]{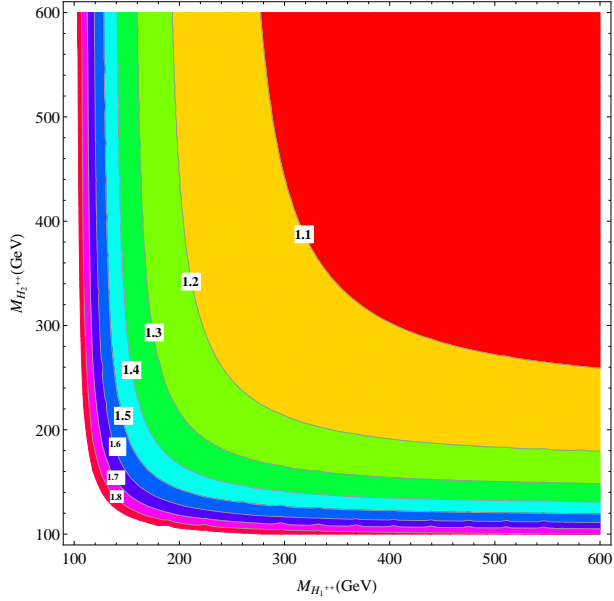}
\end{center}
\caption{$R_{\gamma\gamma}$ with the variation of charged scalar masses.}
\label{fig:Htogammagamma_contour}
\end{figure} 

$A_{1/2}$, $A_{1}$ and $A_0$ are loop functions for fermions, vector bosons and scalars respectively, given as

\begin{eqnarray}
 A_{1/2}(x) &=& 2x^2[x^{-1}+(x^{-1}-1)f(x^{-1})], \\
A_{1}(x) &=& -x^2[2x^{-2}+3x^{-1}+3(2x^{-1}-1)f(x^{-1})], \\
A_{0}(x) &=& -x^2[x^{-1}-f(x^{-1})].
\end{eqnarray}
For the SM-like Higgs mass below threshold, i.e., $ m_{H_0^0} < 2 m_{loop}$  ($m_{loop}$ is a mass of a particle in the loop) 
$f(x) = arcSin^2(\sqrt{x})$.

In Fig.~\ref{fig:Htogammagamma_contour} we present a contour plot to grab the contributions from the charged scalars to $R_{\gamma \gamma}$.
We have assumed $M_{H_1^{++}} = M_{H_1^{+}}$ to reduce number of free parameters.

Experimental observations of the Higgs to di-photon decay normalised to the SM prediction, as pointed out by ATLAS and CMS is given as in \cite{ATLAS:hgg_moriond2013},
\cite{CMS:hgg_moriond2013}:
\begin{eqnarray}
 R_{\gamma\gamma} &=& 1.65 \pm 0.24(stat)^{+ 0.25}_{-0.18}(syst)\; {\rm (ATLAS)\; },\\
&& \nonumber \\ 
 R_{\gamma\gamma} &=& 0.78^{+0.28}_{-0.26}\; {\rm (CMS)\; }.
 \end{eqnarray}

As errors are still very large, it is too early to make any conclusive remarks on these results, especially that tendency seems to be that anomaly systematically approaches 1. However, MLRSM can accommodate wide range of $R_{\gamma \gamma}$ values by the charged Higgs boson effects, for related discussions, see e.g.
in \cite{Dev:2013ff}.

\section{Conclusions and Outlook}

After discovery of the SM-like neutral Higgs boson in July 2012 at the LHC, the next big issue is what is the actual shape of the Higgs potential. Thus a question is asked to reveal the further query regarding  possible gauge symmetry behind the existence of elementary Higgs boson. 
Here we have concentrated on studies connected with LHC potential
discovery of charged Higgs bosons within classical MLRSM which is
already phenomenologically rich enough and worth of separate
investigations.
Though different low energy data and the LHC exclusion plots constrain already $W_2$ and $Z_2$ very much, still the charged scalars can be relatively light. It has been shown which of singly and doubly charged Higgs bosons can be light, in agreement with FCNC limits on neutral Higgs bosons particles, as both charged and neutral scalar sectors are connected through the Higgs potential parameters. They can be produced at the LHC with non-negligible cross sections. 
However, their production cross sections decrease rapidly with their masses, that is why we have undertaken here more detailed and systematic studies  
including the production and decays of charged scalars. We have concentrated on the single and pair production of doubly charged scalars. We have chosen the benchmark points in such a
way that signals connected with doubly charged scalars can dominate over non-standard signals coming from both heavy gauge and neutral Higgs bosons. 
We have analysed the four-lepton and tri-lepton signals at the LHC. 
As a rule of thumb, for all considered processes with doubly charged scalars cross sections are about 1$~fb$ for their masses in vicinity of $400 \div 500$  GeV, which is about the present lowest limit on their masses. If planed integrated luminosity in the next LHC run at $\sqrt{s}=14$ TeV is about 10 times larger than present values, clear signals with four-leptons without missing energy and tri-lepton signals can be detected. It will be an indication for doubly charged scalar effects. These multi lepton final states 
posses very small SM background. We have shown that MLRSM model can give such signals for doubly charged masses up to approximately 600 GeV. 
In our analysis we have used the di-lepton invariant mass and lepton-lepton separation distributions. We also estimate the amount of charge asymmetry in signal as well as background events, and show that this might be a smoking gun feature for future discovery.  The same and opposite sign charged lepton signals have been analysed using proper kinematic cuts and the clear impact of doubly charged scalars are noted carefully. 

Finally, as in the Left-Right symmetric models charged gauge bosons are very heavy, they do not contribute significantly to the Higgs to di-photon process, however, the relatively 
light charged scalars can contribute easily. We have incorporated the impact of the light charged scalars in this process and estimated the strength of this contribution 
over the SM one.

As an outlook, several interesting things can still be done, e.g.
\begin{enumerate}
\item More detailed comparison studies including also lepton spin correlations and their angular distributions with other non-standard models where doubly charged scalars exist (e.g. Higgs Triplet Model \cite{Akeroyd:2012ms}); 
\item Studies of dedicated distributions for processes involving doubly
charged Higgs bosons with both jets and missing energy;
\item Theoretical studies of general Higgs potentials which can realise relatively light charged Higgs bosons keeping at the same time a few TeV scale of neutral Higgs bosons (e.g. \cite{Guadagnoli:2010sd});  
\item To release theoretical assumptions on equality of left and right gauge
boson couplings, diagonal neutrino light-heavy mixings and possible
see-saw scenarios, take into account relations between model parameters
in the Higgs, gauge and neutrino sectors, e.g.  \cite{Czakon:1999ga}.
\end{enumerate}

In summary, we are in a very exciting moment and the next LHC run should be decisive if our scenario with relatively light charged Higgs bosons can be realised. 
Still there is a room for Left-Right gauge symmetry signals discovery at the LHC, including MLRSM doubly charged Higgs bosons effects as long as long as their masses will be well below 1 TeV range ($m_{H_{1/2}^{\pm \pm}} \leq 600$ GeV).

\section*{Acknowledgements}
Work of JC is supported by Department of Science \& Technology, Government of INDIA under the Grant Agreement number IFA12-PH-34 (INSPIRE Faculty Award). Work of MK is supported  by the \'Swider  fellowship.
Work of JG  is supported by the Research Executive Agency (REA) of the European Union under the
Grant Agreement number PITN-GA-2010-264564 (LHCPhenoNet).
Work of RS  is supported by Science and Engineering Research Canada (NSERC). 

\appendix 

\section{Reconciling FCNC effects and large $v_R$ with 
relatively light charged Higgs mass spectrum within MLRSM}  

A scan of potential parameters based on the numerical diagonalisation  and minimisation of the complete MLRSM Higgs potential within our own implementation of the FeynRules package \cite{Christensen:2008py} has been performed. This leads to the Fig.~\ref{higgsspec}. Here, just for illustration, we  discuss it in a simplified form based on approximations discussed in \cite{Gluza:1994ad}.
In MLRSM there is one neutral SM-like Higgs boson having mass proportional to the vacuum expectation value (VEV) $\kappa_1$ ($\sim$ electro-weak breaking scale). The other Higgs bosons are much heavier. A natural mass scale for them is driven by $v_R$ which decides about the  $SU(2)_R\otimes U(1)_{B-L}$ breaking scale. As discussed in the main text of the paper, we assume large $v_R$ ($\sim 8$ TeV), to be consistent with the experimental constraints. 

The minimisation and diagonalisation of the MLRSM Higgs potential have been investigated in  \cite{Gunion:1989in} and explicit correlations among physical and unphysical scalar fields are given in \cite{Gluza:1994ad}. For the sake of completeness, here we have depicted them along with their mass relations considering $\kappa_2=0$:
\begin{itemize}
\item masses
\begin{eqnarray}
 M^2_{H^0_0} &\simeq & 2\kappa^2_1
 \lambda_1, \\ 
 M^2_{H^0_1}&\simeq & \frac{1}{2} \alpha_3 v^2_R, \label{mass1}\\
 M^2_{H^0_2} &\simeq & 2\rho_1 v_R^2, \;\;\;
 M^2_{H^0_3} \simeq  \frac{1}{2} v^2_R
 \left(
 \rho_3 - 2\rho_1
 \right), \\
M^2_{A^0_1} &\simeq& \frac{1}{2} \alpha_3 v^2_R
-2\kappa^2_1
\left(
2\lambda_2-\lambda_3
\right),\\
M^2_{A^0_2} &\simeq &\frac{1}{2}  v^2_R
\left(
\rho_3-2\rho_1
\right),
\\
 M^2_{H^\pm_1} &\simeq& \frac{1}{2}  v^2_R
 \left(
 \rho_3-2\rho_1
 \right)+
 \frac{1}{4}
 \alpha_3 \kappa^2_1 ,\;\;\;
 M^2_{H^\pm_2} \simeq \frac{1}{2} \alpha_3
 \left[
 v^2_R +\frac{1}{2} \kappa^2_1
 \right], \label{masshp} \\
 M^2_{H_1^{\pm \pm}} &\simeq& \frac{1}{2}
 \left[
 v^2_R
 \left(
 \rho_3-2\rho_1
 \right)+
 \alpha_3 \kappa^2_1 
 \right], \;\;\;
 M^2_{H_2^{\pm \pm}} \simeq 2\rho_2 v^2_R+
\frac{1}{2} \alpha_3 \kappa^2_1 . 
\label{mass2}
 \end{eqnarray}
\item relations among physical and unphysical fields (``G" stands for Goldstone modes) 
\begin{eqnarray}
\phi_1^0&\simeq&\frac{1}{\sqrt{2}}\left[
H_0^0+i \tilde{ G_1^0} \right] ,
\\
\phi_2^0 &\simeq& \frac{1}{\sqrt{2}}\left[
H_1^0-i A_1^0 \right], \label{phi20}
\\
\delta_R^0 &=& \frac{1}{\sqrt{2}}\left( H_2^0+iG_2^0 \right),\;\; 
\delta_L^0 = \frac{1}{\sqrt{2}} \left( H_3^0+iA_2^0 \right),\\
\delta_L^+ &=& H_1^+,\;\;\;\delta_R^+ \simeq G_R^+ ,\\
\phi_{1}^+ &\simeq& H_2^+, \;\;\;
\phi_{2}^+ \simeq G_L^+, \\
 \delta_R^{\pm \pm} &=& H_1^{\pm \pm},\;\;\; \delta_L^{\pm \pm} = H_2^{\pm \pm}. 
\end{eqnarray}
\end{itemize}

As masses of quarks are non-degenerate, FCNC effects appear through the $A_0$ part of the following Lagrangian \cite{Duka:1999uc} 
\begin{eqnarray}
 L_{quark-Higgs}(u,d) = &-& \bar{U}
 \left[
 P_L \left( M^u_{diag}B_0^{\ast}+U^{CKM}M^d_{diag}
 U^{CKM\dagger}A_0 \right)
 \right. \nonumber \\
 &+&
 \left.
 P_R \left( M^u_{diag}B_0+U^{CKM}M^d_{diag}
 U^{CKM\dagger}A_0^{\ast} \right)
 \right]
 U, 
 \end{eqnarray}
\\
where $A_0$ is a linear combination of neutral physical
 Higgs and Goldstone fields connected with a bi-doublet $\Phi$
 \cite{Gunion:1989in}, and taking into account Eq.~(\ref{phi20}), we finally have
 
 \begin{eqnarray}
 A_0 &=& \sqrt{2}
 \left(
 \kappa_1 \phi^0_2  
 \right) =
 \left(H^0_1-iA^0_1
 \right).
 \end{eqnarray}
 
 To suppress the effects connected with these fields \cite{Ecker:1983uh,Mohapatra:1983ae,Pospelov:1996fq,Maiezza:2010ic,Chakrabortty:2012pp}, their masses needs to be at least $\sim$ 10 TeV. In our analysis we have kept them to be $\sim$ 15 TeV:
 
 \begin{equation}
 m_{H^0_1},\;m_{A^0_1} > 15\; \rm{ TeV}.
 \label{delfcnc}
 \end{equation}
 
 It can be easily  shown that for defined masses of Higgs bosons, see Eqs.(\ref{mass1}-\ref{mass2}), we can find parameters of the MLRSM Higgs potential within the perturbative limit, and simultaneously satisfy the light charged Higgs bosons and  Eq.~(\ref{delfcnc}). This can be achieved even after keeping three charged Higgs bosons $H_1^{\pm \pm},H_2^{\pm \pm},H_1^{\pm}$ relatively light. For instance, with $v_R=8$ TeV and 
 $\kappa_1=246$ GeV we find the scalar spectrum (in {\rm GeV})
 
\begin{eqnarray}
 M_{H^0_0} &=& 125, \\ 
 M_{H^0_1}&= & 15062,\;\;\;
 M_{H^0_2} =11313, \;\;\;
 M_{H^0_3} = 505, \\
M_{A^0_1} &=& 15066,\\
M_{A^0_2} &= & 505,
\\
 M_{H^\pm_1} &=& 602,\;\;\;
 M_{H^\pm_2} = 15066,  \\
 M_{H_1^{\pm \pm}} &=& 685, \;\;\;
 M_{H_2^{\pm \pm}} = 463, 
 \end{eqnarray} 
  where
  \begin{eqnarray}
  \rho_1 &=& 1,\;\;\;\rho_2=0,\;\;\;\rho_3=2.008,\\
  \lambda_1 &=& 0.13,\;\;\; \lambda_2=0,\;\;\;\lambda_3=1,\\
  \alpha_3&=& 7.09.
  \end{eqnarray}
  
We can see that the remaining fourth charged Higgs boson $H_2^{\pm}$ in MLRSM is naturally very heavy.  To make it light, one needs to go beyond MLRSM and incorporate new terms in the scalar potential which would affect MLRSM Higgs boson masses\footnote{Let us imagine that an additional intermediate energy scale is introduced connected with VEV of an additional $SU(2)_L$ and $SU(2)_R$ singlet scalar field (such scalars give for instance heavy neutrino Majorana mass terms but they decouple from other low energy phenomenological effects).
If this scalar couple to the MLRSM right handed triplet fields, it would modify Eqs.(\ref{masshp}),(\ref{mass2})
but because of its large VEV, mixing of MLRSM Higgs scalars with this state would be negligible, so the effective couplings of MLRSM Higgs bosons, including $H_2^{\pm}$, would stay the same.}. 

\providecommand{\href}[2]{#2}
\addcontentsline{toc}{section}{References}
\bibliographystyle{JHEP}
\bibliography{LRref}

\providecommand{\href}[2]{#2}\begingroup\raggedright\begin{thebibliography}{10}

\bibitem{Mohapatra:1974gc}
R.~Mohapatra and J.~C. Pati, {\it {A Natural Left-Right Symmetry}},  {\em
  Phys.Rev.} {\bf D11} (1975) 2558.

\bibitem{Senjanovic:1975rk}
G.~Senjanovic and R.~N. Mohapatra, {\it {Exact Left-Right Symmetry and
  Spontaneous Violation of Parity}},  {\em Phys.Rev.} {\bf D12} (1975) 1502.

\bibitem{Nemevsek:2011hz}
M.~Nemevsek, F.~Nesti, G.~Senjanovic, and Y.~Zhang, {\it {First Limits on
  Left-Right Symmetry Scale from LHC Data}},  {\em Phys.Rev.} {\bf D83} (2011)
  115014, [\href{http://xxx.lanl.gov/abs/1103.1627}{{\tt arXiv:1103.1627}}].

\bibitem{Ferrari:2000sp}
A.~Ferrari, J.~Collot, M.-L. Andrieux, B.~Belhorma, P.~de~Saintignon, et~al.,
  {\it {Sensitivity study for new gauge bosons and right-handed Majorana
  neutrinos in $p p$ collisions at $s$ = 14-TeV}},  {\em Phys.Rev.} {\bf D62}
  (2000) 013001.

\bibitem{Frank:2011rb}
M.~Frank, A.~Hayreter, and I.~Turan, {\it {Top Quark Pair Production and
  Asymmetry at the Tevatron and LHC in Left-Right Models}},  {\em Phys.Rev.}
  {\bf D84} (2011) 114007, [\href{http://xxx.lanl.gov/abs/1108.0998}{{\tt
  arXiv:1108.0998}}].

\bibitem{Das:2012ii}
S.~Das, F.~Deppisch, O.~Kittel, and J.~Valle, {\it {Heavy Neutrinos and Lepton
  Flavour Violation in Left-Right Symmetric Models at the LHC}},  {\em
  Phys.Rev.} {\bf D86} (2012) 055006,
  [\href{http://xxx.lanl.gov/abs/1206.0256}{{\tt arXiv:1206.0256}}].

\bibitem{Chakrabortty:2012pp}
J.~Chakrabortty, J.~Gluza, R.~Sevillano, and R.~Szafron, {\it {Left-Right
  Symmetry at LHC and Precise 1-Loop Low Energy Data}},  {\em JHEP} {\bf 1207}
  (2012) 038, [\href{http://xxx.lanl.gov/abs/1204.0736}{{\tt
  arXiv:1204.0736}}].

\bibitem{Dev:2013oxa}
P.~S.~B. Dev, C.-H. Lee, and R.~Mohapatra, {\it {Natural TeV-Scale Left-Right
  Seesaw for Neutrinos and Experimental Tests}},
  \href{http://xxx.lanl.gov/abs/1309.0774}{{\tt arXiv:1309.0774}}.

\bibitem{He:2012zp}
X.-G. He and G.~Valencia, {\it {B decays with $\tau$-leptons in non-universal
  left-right models}},  {\em Phys.Rev.} {\bf D87} (2013) 014014,
  [\href{http://xxx.lanl.gov/abs/1211.0348}{{\tt arXiv:1211.0348}}].

\bibitem{Beringer:1900zz}
{\bf Particle Data Group} Collaboration, J.~Beringer et~al., {\it {Review of
  Particle Physics (RPP)}},  {\em Phys.Rev.} {\bf D86} (2012) 010001.

\bibitem{Czakon:1999ga}
M.~Czakon, J.~Gluza, and M.~Zralek, {\it {Low-energy physics and left-right
  symmetry: Bounds on the model parameters}},  {\em Phys.Lett.} {\bf B458}
  (1999) 355--360, [\href{http://xxx.lanl.gov/abs/hep-ph/9904216}{{\tt
  hep-ph/9904216}}].

\bibitem{CMS:kxa}
{\bf CMS} Collaboration, {\it {Search for Narrow Resonances using the Dijet
  Mass Spectrum with 19.6$fb^{-1}$ of pp Collisions at $\sqrt{s}$=8 TeV,
  \rm{CMS-PAS-EXO-12-059}}}, .

\bibitem{CMS:2013qca}
{\bf CMS} Collaboration, {\it {Search for Resonances in the Dilepton Mass
  Distribution in pp Collisions at $\sqrt{s} = 8$ TeV, {CMS-PAS-EXO-12-061}}},
  .

\bibitem{ATLAS:2012qjz}
{\bf ATLAS} Collaboration, {\it {Search for New Phenomena in the Dijet Mass
  Distribution updated using 13.0 $fb^{-1}$ of $pp$ Collisions at $\sqrt{s}=8$
  TeV collected by the ATLAS Detector, \rm{ATLAS-CONF-2012-148,
  ATLAS-COM-CONF-2012-180}}}, .

\bibitem{TheATLAScollaboration:2013iha}
{\bf ATLAS} Collaboration, {\it {Search for $W' \rightarrow t\bar{b}$ in
  proton-proton collisions at a centre-of-mass energy of $\sqrt{s}$ = 8 TeV
  with the ATLAS detector, \rm{ATLAS-CONF-2013-050, ATLAS-COM-CONF-2013-022}}},
  .

\bibitem{ATLAS:2013jma}
{\bf ATLAS} Collaboration, {\it {Search for high-mass dilepton resonances in
  20~$fb^{-1}$ of $pp$ collisions at $\sqrt {s} = 8$~TeV with the ATLAS
  experiment, \rm{ATLAS-CONF-2013-017, ATLAS-COM-CONF-2013-010}}}, .

\bibitem{CMS:2012uaa}
{\bf CMS} Collaboration, {\it {Search for a heavy neutrino and right-handed W
  of the left-right symmetric model in pp collisions at 8 TeV,
  \rm{CMS-PAS-EXO-12-017}}}, .

\bibitem{salo:1558322}
{\bf CMS} Collaboration, C.~O.~U. Vuosalo, {\it {Searches for new physics with
  leptons and/or jets at CMS, \rm{CMS-CR-2013-155, CERN-CMS-CR-2013-155}}}, .

\bibitem{Searches:2001ac}
{\bf LEP Higgs Working Group for Higgs boson searches, ALEPH Collaboration,
  DELPHI Collaboration, L3 Collaboration, OPAL} Collaboration, {\it {Search for
  charged Higgs bosons: Preliminary combined results using LEP data collected
  at energies up to 209-GeV}},
  \href{http://xxx.lanl.gov/abs/hep-ex/0107031}{{\tt hep-ex/0107031}}.

\bibitem{Chatrchyan:2012vca}
{\bf CMS} Collaboration, S.~Chatrchyan et~al., {\it {Search for a light charged
  Higgs boson in top quark decays in $pp$ collisions at $\sqrt{s}=7$ TeV}},
  {\em JHEP} {\bf 1207} (2012) 143,
  [\href{http://xxx.lanl.gov/abs/1205.5736}{{\tt arXiv:1205.5736}}].

\bibitem{ATLAS-CONF-2013-090}
{\bf ATLAS} Collaboration, {\it {Search for charged Higgs bosons in the
  $\tau$+jets final state with pp collision data recorded at $\sqrt s=8$ TeV
  with the ATLAS experiment, {ATLAS-CONF-2013-090}}}, .

\bibitem{CMS:2012kua}
{\bf CMS} Collaboration, {\it {Inclusive search for doubly charged Higgs in
  leptonic final states with the 2011 data at 7 TeV, {CMS-PAS-HIG-12-005}}}, .

\bibitem{ATLAS:2012hi}
{\bf ATLAS} Collaboration, G.~Aad et~al., {\it {Search for doubly-charged Higgs
  bosons in like-sign dilepton final states at $\sqrt{s}=7$ TeV with the ATLAS
  detector}},  {\em Eur.Phys.J.} {\bf C72} (2012) 2244,
  [\href{http://xxx.lanl.gov/abs/1210.5070}{{\tt arXiv:1210.5070}}].

\bibitem{CMS:2012zv}
{\bf CMS} Collaboration, S.~Chatrchyan et~al., {\it {Search for heavy neutrinos
  and W[R] bosons with right-handed couplings in a left-right symmetric model
  in pp collisions at sqrt(s) = 7 TeV}},  {\em Phys.Rev.Lett.} {\bf 109} (2012)
  261802, [\href{http://xxx.lanl.gov/abs/1210.2402}{{\tt arXiv:1210.2402}}].

\bibitem{Aad:2012dm}
{\bf ATLAS} Collaboration, G.~Aad et~al., {\it {ATLAS search for a heavy gauge
  boson decaying to a charged lepton and a neutrino in $pp$ collisions at
  $\sqrt{s}=7$ TeV}},  {\em Eur.Phys.J.} {\bf C72} (2012) 2241,
  [\href{http://xxx.lanl.gov/abs/1209.4446}{{\tt arXiv:1209.4446}}].

\bibitem{Mohapatra:1979ia}
R.~N. Mohapatra and G.~Senjanovic, {\it {Neutrino Mass and Spontaneous Parity
  Violation}},  {\em Phys.Rev.Lett.} {\bf 44} (1980) 912.

\bibitem{Mohapatra:1980yp}
R.~N. Mohapatra and G.~Senjanovic, {\it {Neutrino Masses and Mixings in Gauge
  Models with Spontaneous Parity Violation}},  {\em Phys.Rev.} {\bf D23} (1981)
  165.

\bibitem{Maiezza:2010ic}
A.~Maiezza, M.~Nemevsek, F.~Nesti, and G.~Senjanovic, {\it {Left-Right Symmetry
  at LHC}},  {\em Phys.Rev.} {\bf D82} (2010) 055022,
  [\href{http://xxx.lanl.gov/abs/1005.5160}{{\tt arXiv:1005.5160}}].

\bibitem{Tello:2010am}
V.~Tello, M.~Nemevsek, F.~Nesti, G.~Senjanovic, and F.~Vissani, {\it
  {Left-Right Symmetry: from LHC to Neutrinoless Double Beta Decay}},  {\em
  Phys.Rev.Lett.} {\bf 106} (2011) 151801,
  [\href{http://xxx.lanl.gov/abs/1011.3522}{{\tt arXiv:1011.3522}}].

\bibitem{Nemevsek:2011aa}
M.~Nemevsek, F.~Nesti, G.~Senjanovic, and V.~Tello, {\it {Neutrinoless Double
  Beta Decay: Low Left-Right Symmetry Scale?}},
  \href{http://xxx.lanl.gov/abs/1112.3061}{{\tt arXiv:1112.3061}}.

\bibitem{Chakrabortty:2012mh}
J.~Chakrabortty, H.~Z. Devi, S.~Goswami, and S.~Patra, {\it {Neutrinoless
  double-$\beta$ decay in TeV scale Left-Right symmetric models}},  {\em JHEP}
  {\bf 1208} (2012) 008, [\href{http://xxx.lanl.gov/abs/1204.2527}{{\tt
  arXiv:1204.2527}}].

\bibitem{Dev:2013vxa}
P.~Bhupal~Dev, S.~Goswami, M.~Mitra, and W.~Rodejohann, {\it {Constraining
  Neutrino Mass from Neutrinoless Double Beta Decay}},
  \href{http://xxx.lanl.gov/abs/1305.0056}{{\tt arXiv:1305.0056}}.

\bibitem{Shaban:1992he}
N.~Shaban and W.~J. Stirling, {\it {Minimal left-right symmetry and SO(10)
  grand unification using LEP coupling constant measurements}},  {\em
  Phys.Lett.} {\bf B291} (1992) 281--287.

\bibitem{Chakrabortty:2009xm}
J.~Chakrabortty and A.~Raychaudhuri, {\it {GUTs with dim-5 interactions: Gauge
  Unification and Intermediate Scales}},  {\em Phys.Rev.} {\bf D81} (2010)
  055004, [\href{http://xxx.lanl.gov/abs/0909.3905}{{\tt arXiv:0909.3905}}].

\bibitem{Gunion:1989in}
J.~Gunion, J.~Grifols, A.~Mendez, B.~Kayser, and F.~I. Olness, {\it {Higgs
  Bosons in Left-Right Symmetric Models}},  {\em Phys.Rev.} {\bf D40} (1989)
  1546.

\bibitem{Duka:1999uc}
P.~Duka, J.~Gluza, and M.~Zralek, {\it {Quantization and renormalization of the
  manifest left-right symmetric model of electroweak interactions}},  {\em
  Annals Phys.} {\bf 280} (2000) 336--408,
  [\href{http://xxx.lanl.gov/abs/hep-ph/9910279}{{\tt hep-ph/9910279}}].

\bibitem{Czakon:2002wm}
M.~Czakon, J.~Gluza, and J.~Hejczyk, {\it {Muon decay to one loop order in the
  left-right symmetric model}},  {\em Nucl.Phys.} {\bf B642} (2002) 157--172,
  [\href{http://xxx.lanl.gov/abs/hep-ph/0205303}{{\tt hep-ph/0205303}}].

\bibitem{Gluza:2002vs}
J.~Gluza, {\it {On teraelectronvolt Majorana neutrinos}},  {\em Acta
  Phys.Polon.} {\bf B33} (2002) 1735--1746,
  [\href{http://xxx.lanl.gov/abs/hep-ph/0201002}{{\tt hep-ph/0201002}}].

\bibitem{Gluza:1993gf}
J.~Gluza and M.~Zralek, {\it {Neutrino production in e+ e- collisions in a
  left-right symmetric model}},  {\em Phys.Rev.} {\bf D48} (1993) 5093--5105.

\bibitem{Mohapatra:1988tm}
R.~N. Mohapatra and S.~Nussinov, {\it {Constraints on decaying right-handed
  Majorana neutrinos from SN1987a observations}},  {\em Phys.Rev.} {\bf D39}
  (1989) 1378--1385.

\bibitem{Czakon:1999ue}
M.~Czakon, M.~Zralek, and J.~Gluza, {\it {Left-right symmetry and heavy
  particle quantum effects}},  {\em Nucl.Phys.} {\bf B573} (2000) 57--74,
  [\href{http://xxx.lanl.gov/abs/hep-ph/9906356}{{\tt hep-ph/9906356}}].

\bibitem{Belyaev:2012qa}
A.~Belyaev, N.~D. Christensen, and A.~Pukhov, {\it {CalcHEP 3.4 for collider
  physics within and beyond the Standard Model}},  {\em Comput.Phys.Commun.}
  {\bf 184} (2013) 1729--1769, [\href{http://xxx.lanl.gov/abs/1207.6082}{{\tt
  arXiv:1207.6082}}].

\bibitem{Mangano:2002ea}
M.~L. Mangano, M.~Moretti, F.~Piccinini, R.~Pittau, and A.~D. Polosa, {\it
  {ALPGEN, a generator for hard multiparton processes in hadronic collisions}},
   {\em JHEP} {\bf 0307} (2003) 001,
  [\href{http://xxx.lanl.gov/abs/hep-ph/0206293}{{\tt hep-ph/0206293}}].

\bibitem{Sjostrand:2006za}
T.~Sjostrand, S.~Mrenna, and P.~Z. Skands, {\it {PYTHIA 6.4 Physics and
  Manual}},  {\em JHEP} {\bf 0605} (2006) 026,
  [\href{http://xxx.lanl.gov/abs/hep-ph/0603175}{{\tt hep-ph/0603175}}].

\bibitem{Alwall:2011uj}
J.~Alwall, M.~Herquet, F.~Maltoni, O.~Mattelaer, and T.~Stelzer, {\it {MadGraph
  5 : Going Beyond}},  {\em JHEP} {\bf 1106} (2011) 128,
  [\href{http://xxx.lanl.gov/abs/1106.0522}{{\tt arXiv:1106.0522}}].

\bibitem{Christensen:2008py}
N.~D. Christensen and C.~Duhr, {\it {FeynRules - Feynman rules made easy}},
  {\em Comput.Phys.Commun.} {\bf 180} (2009) 1614--1641,
  [\href{http://xxx.lanl.gov/abs/0806.4194}{{\tt arXiv:0806.4194}}].

\bibitem{Degrande:2011ua}
C.~Degrande, C.~Duhr, B.~Fuks, D.~Grellscheid, O.~Mattelaer, et~al., {\it {UFO
  - The Universal FeynRules Output}},  {\em Comput.Phys.Commun.} {\bf 183}
  (2012) 1201--1214, [\href{http://xxx.lanl.gov/abs/1108.2040}{{\tt
  arXiv:1108.2040}}].

\bibitem{Guadagnoli:2010sd}
D.~Guadagnoli and R.~N. Mohapatra, {\it {TeV Scale Left Right Symmetry and
  Flavor Changing Neutral Higgs Effects}},  {\em Phys.Lett.} {\bf B694} (2011)
  386--392, [\href{http://xxx.lanl.gov/abs/1008.1074}{{\tt arXiv:1008.1074}}].

\bibitem{delAguila:2008pw}
F.~del Aguila, J.~de~Blas, and M.~Perez-Victoria, {\it {Effects of new leptons
  in Electroweak Precision Data}},  {\em Phys.Rev.} {\bf D78} (2008) 013010,
  [\href{http://xxx.lanl.gov/abs/0803.4008}{{\tt arXiv:0803.4008}}].

\bibitem{Gluza:1997ts}
J.~Gluza, J.~Maalampi, M.~Raidal, and M.~Zralek, {\it {Heavy neutrino mixing
  and single production at linear collider}},  {\em Phys.Lett.} {\bf B407}
  (1997) 45--52, [\href{http://xxx.lanl.gov/abs/hep-ph/9703215}{{\tt
  hep-ph/9703215}}].

\bibitem{Gluza:1995ix}
J.~Gluza and M.~Zralek, {\it {On possibility of detecting the e-e- to W-W-
  process in the standard model}},  {\em Phys.Lett.} {\bf B362} (1995)
  148--154, [\href{http://xxx.lanl.gov/abs/hep-ph/9507269}{{\tt
  hep-ph/9507269}}].

\bibitem{Gluza:1995ky}
J.~Gluza and M.~Zralek, {\it {Inverse neutrinoless double beta decay in gauge
  theories with CP violation}},  {\em Phys.Rev.} {\bf D52} (1995) 6238--6248,
  [\href{http://xxx.lanl.gov/abs/hep-ph/9502284}{{\tt hep-ph/9502284}}].

\bibitem{Keung:1983uu}
W.-Y. Keung and G.~Senjanovic, {\it {Majorana Neutrinos and the Production of
  the Right-handed Charged Gauge Boson}},  {\em Phys.Rev.Lett.} {\bf 50} (1983)
  1427.

\bibitem{Chen:2013foz}
C.-Y. Chen, P.~S.~B. Dev, and R.~Mohapatra, {\it {Probing Heavy-Light Neutrino
  Mixing in Left-Right Seesaw Models at the LHC}},  {\em Phys.Rev.} {\bf D88}
  (2013) 033014, [\href{http://xxx.lanl.gov/abs/1306.2342}{{\tt
  arXiv:1306.2342}}].

\bibitem{Huitu:1996su}
K.~Huitu, J.~Maalampi, A.~Pietila, and M.~Raidal, {\it {Doubly charged Higgs at
  LHC}},  {\em Nucl.Phys.} {\bf B487} (1997) 27--42,
  [\href{http://xxx.lanl.gov/abs/hep-ph/9606311}{{\tt hep-ph/9606311}}].

\bibitem{PhysRevD.40.1521}
M.~L. Swartz, {\it Limits on doubly charged higgs bosons and lepton-flavor
  violation},  {\em Phys. Rev. D} {\bf 40} (Sep, 1989) 1521--1528.

\bibitem{Rizzo:1981xx}
T.~G. Rizzo, {\it {Doubly Charged Higgs Bosons and Lepton Number Violating
  Processes}},  {\em Phys.Rev.} {\bf D25} (1982) 1355--1364.

\bibitem{Maalampi:2002vx}
J.~Maalampi and N.~Romanenko, {\it {Single production of doubly charged Higgs
  bosons at hadron colliders}},  {\em Phys.Lett.} {\bf B532} (2002) 202--208,
  [\href{http://xxx.lanl.gov/abs/hep-ph/0201196}{{\tt hep-ph/0201196}}].

\bibitem{delAguila:2008cj}
F.~del Aguila and J.~Aguilar-Saavedra, {\it {Distinguishing seesaw models at
  LHC with multi-lepton signals}},  {\em Nucl.Phys.} {\bf B813} (2009) 22--90,
  [\href{http://xxx.lanl.gov/abs/0808.2468}{{\tt arXiv:0808.2468}}].

\bibitem{Melfo:2011nx}
A.~Melfo, M.~Nemevsek, F.~Nesti, G.~Senjanovic, and Y.~Zhang, {\it {Type II
  Seesaw at LHC: The Roadmap}},  {\em Phys.Rev.} {\bf D85} (2012) 055018,
  [\href{http://xxx.lanl.gov/abs/1108.4416}{{\tt arXiv:1108.4416}}].

\bibitem{delAguila:2013yaa}
F.~del Aguila, M.~Chala, A.~Santamaria, and J.~Wudka, {\it {Discriminating
  between lepton number violating scalars using events with four and three
  charged leptons at the LHC}},  {\em Phys.Lett.} {\bf B725} (2013) 310--315,
  [\href{http://xxx.lanl.gov/abs/1305.3904}{{\tt arXiv:1305.3904}}].

\bibitem{Babu-Ayon}
K.~Babu, A.~Patra, and S.~K. Rai, {\it {New Signals for Doubly-Charged Scalars
  and Fermions at the Large Hadron Collider}},
  \href{http://xxx.lanl.gov/abs/1306.2066}{{\tt arXiv:1306.2066}}.

\bibitem{delAguila:2013mia}
F.~del Aguila and M.~Chala, {\it {LHC bounds on Lepton Number Violation
  mediated by doubly and singly-charged scalars}},
  \href{http://xxx.lanl.gov/abs/1311.1510}{{\tt arXiv:1311.1510}}.

\bibitem{Pumplin:2002vw}
J.~Pumplin, D.~Stump, J.~Huston, H.~Lai, P.~M. Nadolsky, et~al., {\it {New
  generation of parton distributions with uncertainties from global QCD
  analysis}},  {\em JHEP} {\bf 0207} (2002) 012,
  [\href{http://xxx.lanl.gov/abs/hep-ph/0201195}{{\tt hep-ph/0201195}}].

\bibitem{gb-jd-sg-pk}
G.~Bambhaniya, J.~Chakrabortty, S.~Goswami, and P.~Konar, {\it {Generation of
  Neutrino mass from new physics at TeV scale and Multi-lepton Signatures at
  the LHC}},  \href{http://xxx.lanl.gov/abs/1305.2795}{{\tt arXiv:1305.2795}}.

\bibitem{Kom:2010mv}
C.-H. Kom and W.~J. Stirling, {\it {Charge asymmetry in W + jets production at
  the LHC}},  {\em Eur.Phys.J.} {\bf C69} (2010) 67--73,
  [\href{http://xxx.lanl.gov/abs/1004.3404}{{\tt arXiv:1004.3404}}].

\bibitem{PhysRevD.40.1722}
R.~Martinez, M.~A. Perez, and J.~J. Toscano, {\it Two-photon decay width of the
  higgs boson in left-right-symmetric theories},  {\em Phys. Rev. D} {\bf 40}
  (Sep, 1989) 1722--1724.

\bibitem{Picek:2012ei}
I.~Picek and B.~Radovcic, {\it {Enhancement of $h \to \gamma \gamma$ by
  seesaw-motivated exotic scalars}},  {\em Phys.Lett.} {\bf B719} (2013)
  404--408, [\href{http://xxx.lanl.gov/abs/1210.6449}{{\tt arXiv:1210.6449}}].

\bibitem{Carena:2012xa}
M.~Carena, I.~Low, and C.~E. Wagner, {\it {Implications of a Modified Higgs to
  Diphoton Decay Width}},  {\em JHEP} {\bf 1208} (2012) 060,
  [\href{http://xxx.lanl.gov/abs/1206.1082}{{\tt arXiv:1206.1082}}].

\bibitem{ATLAS:hgg_moriond2013}
{\bf ATLAS} Collaboration, {\it {Measurements of the properties of the
  Higgs-like boson in the two photon decay channel with the ATLAS detector
  using 25 $\mathrm{fb}^{-1}$ of proton-proton collision data}},
  \href{http://xxx.lanl.gov/abs/ATLAS-CONF-2013-012,
  ATLAS-COM-CONF-2013-015}{{\tt ATLAS-CONF-2013-012, ATLAS-COM-CONF-2013-015}}.

\bibitem{CMS:hgg_moriond2013}
{\bf CMS} Collaboration, {\it {Updated measurements of the Higgs boson at 125
  GeV in the two photon decay channel}},
  \href{http://xxx.lanl.gov/abs/CMS-PAS-HIG-13-001}{{\tt CMS-PAS-HIG-13-001}}.

\bibitem{Dev:2013ff}
P.~Bhupal~Dev, D.~K. Ghosh, N.~Okada, and I.~Saha, {\it {125 GeV Higgs Boson
  and the Type-II Seesaw Model}},  {\em JHEP} {\bf 1303} (2013) 150,
  [\href{http://xxx.lanl.gov/abs/1301.3453}{{\tt arXiv:1301.3453}}].

\bibitem{Akeroyd:2012ms}
A.~Akeroyd and S.~Moretti, {\it {Enhancement of H to gamma gamma from doubly
  charged scalars in the Higgs Triplet Model}},  {\em Phys.Rev.} {\bf D86}
  (2012) 035015, [\href{http://xxx.lanl.gov/abs/1206.0535}{{\tt
  arXiv:1206.0535}}].

\bibitem{Gluza:1994ad}
J.~Gluza and M.~Zralek, {\it {Higgs boson contributions to neutrino production
  in e- e+ collisions in a left-right symmetric model}},  {\em Phys.Rev.} {\bf
  D51} (1995) 4695--4706, [\href{http://xxx.lanl.gov/abs/hep-ph/9409225}{{\tt
  hep-ph/9409225}}].

\bibitem{Ecker:1983uh}
G.~Ecker, W.~Grimus, and H.~Neufeld, {\it {Higgs Induced Flavor Changing
  Neutral Interactions in SU(2)-l X SU(2)-r X U(1)}},  {\em Phys.Lett.} {\bf
  B127} (1983) 365.

\bibitem{Mohapatra:1983ae}
R.~N. Mohapatra, G.~Senjanovic, and M.~D. Tran, {\it {Strangeness Changing
  Processes and the Limit on the Right-handed Gauge Boson Mass}},  {\em
  Phys.Rev.} {\bf D28} (1983) 546.

\bibitem{Pospelov:1996fq}
M.~Pospelov, {\it {FCNC in left-right symmetric theories and constraints on the
  right-handed scale}},  {\em Phys.Rev.} {\bf D56} (1997) 259--264,
  [\href{http://xxx.lanl.gov/abs/hep-ph/9611422}{{\tt hep-ph/9611422}}].

\end{thebibliography}\endgroup
\end{document}